\documentclass[]{aastex631}

\usepackage{makecell}
\usepackage{color}
\usepackage{textcomp}
\usepackage{bm}
\usepackage{hyperref}
\hypersetup{hidelinks,colorlinks=true,allcolors=blue,breaklinks=true}

\shortauthors{Liu et al.}

\graphicspath{{./}{figures/}}

\begin{document}

\title{Convection and Clouds under Different Planetary Gravities Simulated by a Small-domain Cloud-resolving Model}

\correspondingauthor{Jun Yang}
\email{junyang@pku.edu.cn}

\author[0000-0001-8981-1798]{Jiachen Liu}
\affiliation{Dept. of Atmospheric and Oceanic Sciences, School of Physics, Peking University, Beijing 100871, People's Republic of China}

\author[0000-0001-6031-2485]{Jun Yang}
\affiliation{Dept. of Atmospheric and Oceanic Sciences, School of Physics, Peking University, Beijing 100871, People's Republic of China}

\author[0000-0001-6194-760X]{Yixiao Zhang}
\affiliation{Dept. of Atmospheric and Oceanic Sciences, School of Physics, Peking University, Beijing 100871, People's Republic of China}

\author[0000-0002-7422-3317]{Zhihong Tan}
\affiliation{Cooperative Institute for Modeling the Earth System, Princeton University and NOAA Geophysical Fluid Dynamics Laboratory, Princeton, NJ 08540, USA}

\begin{abstract}

In this study, we employ a cloud-resolving model (CRM) to investigate how gravity influences convection and clouds in a small-domain (96 km by 96 km) radiative-convective equilibrium (RCE). Our experiments are performed with a horizontal grid spacing of 1 km, which can resolve large (\textgreater 1 km$^2$) convective cells. We find that under a given stellar flux, sea surface temperature increases with decreasing gravity. This is because a lower-gravity planet has larger water vapor content and more clouds, resulting in a larger clear-sky greenhouse effect and a stronger cloud warming effect in the small domain. By increasing stellar flux under different gravity values, we find that the convection shifts from a quasi-steady state to an oscillatory state. In the oscillatory state, there are convection cycles with a period of several days, comprised of a short wet phase with intense surface precipitation and a dry phase with no surface precipitation. When convection shifts to the oscillatory state, water vapor content and high-level cloud fraction increase substantially, resulting in rapid warming. After the transition to the oscillatory state, the cloud net positive radiative effect decreases with increasing stellar flux, which indicates a stabilizing climate effect. In the quasi-steady state, the atmospheric absorption features of CO$_2$ are more detectable on lower-gravity planets because of their larger atmospheric heights. While in the oscillatory state, the high-level clouds mute almost all the absorption features, making the atmospheric components hard to be characterized.

\end{abstract}


\section{Introduction}\label{intro}

Exploring potentially habitable planets beyond the solar system is the main goal of exoplanet missions. 
Planetary climate and the range of the habitable zone depend on various factors, including the stellar flux, stellar spectrum, planetary mass, radius, gravity, rotation rate, obliquity, atmospheric mass, and composition \citep[e.g.,][]{kopparapu2013habitable,turbet2016habitability,salameh2018role,yang2019planetary,colose2019enhanced,yang2019effects,madden2020high}. Among them, planetary gravity is easy to estimate based on the measurements of planetary mass and radius using the two methods of transit and radial velocity \citep{weiss2014mass}. Understanding the effects of gravity on planetary climate and the habitable zone can help scientists better target potentially habitable exoplanets. 

Several studies have investigated the effects of planetary gravity. \citet{pierrehumbert2010principles} applied a 1D radiative-convective model to explore the influences of gravity under a water-vapor-only atmosphere. He illustrated that the limiting outgoing longwave radiation (OLR) is smaller on lower-gravity planets, so planets with smaller values of gravity require smaller stellar flux to enter the runaway greenhouse state. This is because, under a lower gravity, less water vapor pressure is needed to obtain a given water vapor mass to enter the runaway greenhouse state.
\citet{kopparapu2014habitable} also suggested that the limiting OLR is smaller on lower-gravity planets. \citet{yang2019effects} employed a 3D GCM (ExoCAM) to demonstrate that a smaller gravity value makes the inner edge of the habitable zone of tidally locked planets move further away from the host stars, consistent with 1D models. Also with 3D GCMs, \citet{yang2019planetary} and \citet{thomson2019effects} stated that because lower-gravity planets have larger water column depths, they have stronger greenhouse effects and warmer climates. In particular, \citet{thomson2019effects} demonstrated that in a dry atmosphere, with hydrostatic equilibrium being assumed and the radiative forcing unaltered, the effect of gravity can be simply scaled: if gravity varies by a factor of $\alpha$, the vertical coordinate \textit{z} changes by a factor of 1/$\alpha$, with atmospheric circulation being unchanged. However, in an atmosphere with condensible components such as water vapor, varying gravity will influence both climate and atmospheric circulation.

Based on the studies mentioned above, it is clear that gravity can significantly influence vapor content. However, the effects of gravity on clouds and convection which are closely related to the vapor field were poorly estimated. Convection and clouds were not considered or well represented in previous studies using 1D radiative-convective models. 
With the coarse resolution of hundreds of kilometers, GCMs can not simulate processes of convection on scales of meters to kilometers, so they have to use cumulus parameterization schemes. The cumulus parameterization schemes involve many empirical equations and parameters based on Earth, raising the question of whether they can apply to different values of gravity. 
In this study, we overcome this limitation by employing a cloud-resolving model (CRM) with a horizontal grid spacing of 1 km, which is fine enough to resolve large (\textgreater 1 km$^2$) convective cells, such as shallow cumulus, deep convective plumes, and tropical anvil clouds \citep[e.g.,][]{ResolvedSnowballEarthClouds,satoh2019global}.

Clouds play vital roles in planetary climate, by regulating the amount of radiation in the atmosphere and at the surface \citep{stephens2005cloud}. Several previous works have employed CRMs in the study of exoplanets and planetary climate \citep[e.g.,][]{zhang2017surface,Sergeev_2020,Tan_2021,seeley2021episodic,lefevre20213d,Lef_vre_2022,Song_2022}. \citet{zhang2017surface} used a mesoscale model with gird spacing of 3 km in a 1000 by 1000 km domain and found that small-scale updrafts lead to non-uniform distributions of water vapor, clouds, surface shortwave radiation, and surface temperature near the substellar point on tidally locked terrestrial planets around M stars. This study suggested that a more realistic description of convection and clouds with a CRM can provide a clearer picture of cloud effects on the climate of exoplanets around M stars. \citet{Sergeev_2020} investigated the dependence of the climate of tidally locked terrestrial exoplanets on parameterized convection schemes and explicit substellar convection. They found that with explicit substellar convection, the surface temperature contrast between dayside and nightside is higher than that with convection parameterization schemes, which suggested that a more realistic description of convection and clouds can help improve the exoplanet simulations. 
\citet{seeley2021episodic} employed a cloud-resolving model in a 72 by 72 km small domain and found that lower-tropospheric radiative heating causes convection to shift from a quasi-steady regime to an oscillatory regime, in which precipitation occurs as an intense outburst separated by a several-day dry period. This study featured a novel form of temporal convective self-organization. This convective self-organization was not seen in previous GCM simulations, because their horizontal resolutions are too coarse for small-scale processes.

In this study, we employ a CRM in the frame of radiative-convective equilibrium (RCE) coupled to a slab ocean to estimate the effects of gravity on the planetary climate. RCE is an idealization of the tropical atmosphere, in which radiative cooling is mainly balanced by convective heating. It is widely used in studying the essential interactions between convection and radiative transfer on Earth. We perform our experiments in a small domain, 96 by 96 km, similar to \citet{seeley2021episodic}. By explicitly resolving convection, we can study the interaction between radiative heating, convective-scale air motions, and clouds. To carefully estimate convection and clouds under different gravity values and meanwhile save computation resources, we only perform our experiments in a small domain in this study. The oscillatory convection found in \citet{seeley2021episodic} also occurs in our experiments with high stellar fluxes and small gravity values. Different from \citet{seeley2021episodic}, we focus more on the impacts of gravity on the planetary climate and also analyze the effects of gravity under lower surface temperatures without the oscillatory convection. 

The outline of the article is as follows. In Section \ref{methods}, we introduce the CRM used in our study and the experimental designs. In Sections \ref{subsec: fixed} and \ref{subsec: increasing}, we show the effects of gravity on the planetary climate under a fixed stellar flux and with increasing stellar flux. Sensitivities to the fall speeds of precipitation droplets and the threshold temperature between cloud liquid water and cloud ice water are discussed in Section \ref{sec: sensitivity}. The analysis of transmission spectra is shown in Section \ref{sec: transmission}. Finally, conclusions and discussions are demonstrated in Section \ref{conclusions}. We find that under a given stellar flux, sea surface temperature increases with decreasing gravity. With increasing stellar flux under different gravity values, we find that the convection shifts from a quasi-steady state to an oscillatory state. Climate feedbacks change significantly before, at, and after the transition to the oscillatory state. With the same sea surface temperature, lower-gravity planets can hold more water vapor, so they shift to the oscillatory state under a lower sea surface temperature. In the quasi-steady state, atmospheric components, especially CO$_2$, are more detectable on lower gravity planets. While in the oscillatory state, clouds mute almost all the absorption features of CO$_2$.

\section{Model Descriptions and Experimental Designs} \label{methods}

\begin{deluxetable}{cccccc}
\tablenum{1}
\tablecaption{Summary of the Numerical Experiments in this Study.}\label{table 1}
\tablehead{\colhead{Experiment} & \colhead{Gravity} & 
\colhead{Model-top}&
\colhead{Vertical levels} & 
\colhead{Stellar flux} & 
\colhead{Notes}
\\
\colhead{} &
\colhead{(g$_{e}$)} & 
\colhead{(km)} & 
\colhead{} & 
\colhead{(W\,m$^{-2}$)}}
\startdata
& 0.38 & 150 & 165 & 230 & \\
& 0.5 & 90 & 105 &  230 & \\
under a fixed & 0.75 & 50 & 65 & 230 & \\
stellar flux  & 1.0 & 50 & 65 & 230 &\\
& 1.25 & 35 & 40 & 230 & \\
& 1.5 & 25 & 30 & 230 & \\
\hline
& 0.38 & 70 & 85 & 270 &\\
without cloud & 0.5 & 50 & 65 & 270 & \\
radiative effects& 0.75 & 50 & 65 & 270 & \\
& 1.0 & 50 & 65 & 270 &\\
\hline
& 0.5 & 90 & 105 & 230 &
\\M star spectrum & 1.0 & 50 & 65 & 230 & 
\\& 1.5 & 25 & 30 & 230 &\\
\hline
fixed & 0.5 & 90 & 105 & 230 &
\\
surface pressure & 1.5 & 25 & 30 & 230 &\\
\hline
& 0.38 & 110 or 150  & 125 or 165 & 200--250 & \\
under increasing & 0.5 & 90 & 105 & 230--270 & \\
stellar flux & 0.75 & 50 & 65 & 230--270 & \\
at 10--W\,m$^{-2}$ & 1.0 & 50 & 65 & 230--290 &\\
increments & 1.25 & 35 & 40 & 230--300 & \\
& 1.5 & 25 & 30 & 230--300 & \\
\hline
sensitivity & 0.5 & 90 & 105 & 230 & not changed\\
to the & 0.5 & 90 & 105 & 250 & SAM's default
 \\
fall speeds & 1.5 & 25 & 30 & 230 & fall speeds \\ 
\hline
\makecell{sensitivity to \\the threshold\\ temperature}  & 0.38 & 150 & 165 & 230 & \makecell{switch the threshold\\ temperature from \\253 K to 233 K} \\
\enddata
\tablecomments{We use g$_{e}$ to represent Earth's gravity value: 9.81 m s$^{-2}$. The fall speeds of precipitation droplets are multiplied by 0.5$^{th}$ power of the ratio of the simulated gravity to Earth's gravity in the control runs. The model-top height of the cases of 0.38g$_e$ with 200, 210, 220 W\,m$^{-2}$ is set to be 110 km, and the model-top height of the cases of 0.38g$_e$ with 230, 240, 250 W\,m$^{-2}$ is set to be 150 km, because convection extends to higher altitudes in the latter cases.}
\end{deluxetable}

The model employed in this study is the System for Atmospheric Modeling (SAM) version 6.11.6, documented by \citet{khairoutdinov2003cloud}. SAM uses the anelastic dynamical core, and the scalar advection is formulated with positive and monotonic schemes of \citet{smolarkiewicz1990multidimensional}. 
The partitioning of the diagnosed water variables into hydrometeor mixing ratios is done as a function of temperature. We use a single-moment microphysics scheme \citep{khairoutdinov2006high}. The calculation of shortwave and longwave radiation uses the radiative transfer model adapted from the Community Atmospheric
Model version 3.0 \citep[CAM3;][]{collins2004description,collins2006formulation}. We do not use the Rapid Radiative Transfer Model (RRTM) even though it is well validated for modern Earth. This is because RRTM uses different look-up tables for water vapor below and above 100 hPa. If water vapor content is large near 100 hPa (which is the case in our lower-gravity simulations),  RRTM produces unphysical and discontinuous heating rates \citep[see Extended Figure 1 in][and also Figure \ref{compare} below]{seeley2021episodic}. 

A wide range of gravity values are examined, 0.38 to 1.5 times Earth's value. Below, we will use g$_e$ to represent Earth’s gravity, 9.81 m\,s$^{-2}$. The lower limit, 0.38g$_e$, is the gravity value of Mars. We do not examine gravity values smaller than 0.38g$_e$, although moons such as Titan (0.14g$_e$) can also maintain long-lived, substantial atmospheres \citep{lammer2014origin,heller2014formation}. This is because the model is easy to crash with an extremely low gravity value, in which the radiative heating rate is very large (reaches $\sim$100 K\,day$^{-1}$) in high altitudes (figure not shown). The upper bound is approximately set to be the upper limit of the gravity of super-Earths \citep{fulton2017california}. 

All experiments are three-dimensional and employed in a small domain, 96 km in \textit{x} direction by 96 km in \textit{y} direction, with a horizontal grid spacing of 1 km. Due to computation resource limitation, we do not perform experiments in a larger domain. Different runs have different numbers of levels and model-top heights, but the same vertical grid spacing: the first level is at 50 m, then grid spacing gradually increases from 100 m at the second level to 1 km at 10 km; above 10 km, the spacing is set to be constant, 1 km. We have tested different vertical resolutions, and find that the results do not change significantly. To simulate deep convection, the model-top heights of lower-gravity experiments should be much larger than those of higher-gravity. This is because lower-gravity planets have smaller lapse rates (for example, the dry adiabatic lapse rate is $g/c_{p}$, where $g$ is gravity and $c_{p}$ is the specific heat of air) and meanwhile  larger scale heights  \citep[$H=R\left\langle T\right\rangle/g$, where $R$ is the air constant and $\left\langle T\right \rangle$ is the vertical-mean air temperature;][]{hartmann1994global}. The model-top heights of the experiments of 0.38g$_e$, 0.5g$_e$, 0.75g$_e$, 1.0g$_e$, 1.25g$_e$, and 1.5g$_e$ are set to be 110 or 150, 90, 50, 50, 35, and 25 km respectively (Table \ref{table 1}). Boundary conditions are periodic in both \textit{x} and \textit{y} directions. The top boundary is not periodic, so a sponge layer is set in the upper 1/10 levels of the model to reduce gravity wave reflection, in which Newtonian damping of prognostic variables is done.

Atmospheric composition is set to be Earth-like, a N$_{2}$-dominated atmosphere with CO$_{2}$ and O$_{2}$. CO$_{2}$ mixing ratio is 355.5 ppmv. We set the column air mass to Earth's value, 1.0\,$\times$\,10$^{4}$ kg\,m$^{-2}$, and assume that when the planetary gravity is modified, the column air mass does not change. So, surface pressure varies proportionally to planetary gravity. We use the solar spectrum in most experiments, and experiments with an M star spectrum are 
tested to expand our results to exoplanets around M stars. The stellar flux is made perpetual and homogeneous without diurnal or seasonal cycle, with a zenith angle of 0$^\circ$. The model is coupled to a 2 m slab ocean, so the sea surface is actively coupled to the atmosphere. The ice-free ocean albedo is a function of the solar zenith angle in SAM. Since we set the solar zenith angle to zero, the ocean albedo is a constant value of 0.024.
Coriolis force and large-scale forcing are not considered. 

The cloud particle radii are parameterized based on CAM3 in our simulations \citep{collins2004description,collins2006formulation}, in which cloud liquid radius is set to 14 \textmu m and cloud ice radii are prescribed as a function of temperature only. We assume the cloud particle radii do not change with gravity for simplicity. SAM uses three constant fall speeds for rain, snow, and graupel, respectively. The fall speeds of precipitation droplets should change with varying gravity \citep{bohm1989general}. Thus, we multiply the fall speeds of precipitation droplets in SAM by 0.5$^{th}$ power of the ratio of the simulated gravity to Earth's gravity according to \citet{khvorostyanov2002terminal}. For example, for the cases of 0.38g$_e$ and 1.5g$_e$, we multiply the fall speeds of precipitation droplets by 0.62 and 1.22, respectively.

We first investigate the effects of planetary gravity under a fixed stellar flux, 230 W\,m$^{-2}$. Then, we estimate how gravity affects planetary climate under increasing stellar flux at a 10--W\,m$^{-2}$ increment. For the cases of 0.5g$_e$, 0.75g$_e$, 1.0g$_e$, 1.25g$_e$, and 1.5g$_e$, we increase the stellar flux from 230 W\,m$^{-2}$ to 300 W\,m$^{-2}$ or until the model crashes under high surface temperatures. For the cases of 0.38g$_e$, convection has entered the oscillatory state with 230 W\,m$^{-2}$, so we start to increase the stellar flux from 200 W\,m$^{-2}$ to simulate the quasi-steady state. Then, we perform two groups of sensitivity tests on the fall speeds of precipitation droplets and the threshold temperature between cloud liquid water and cloud ice water. After that, we estimate the effects of gravity on the transmission spectra using the Planetary Spectrum Generator \citep[PSG;][]{VILLANUEVA201886}. More detailed settings are shown in Section \ref{sec: transmission}.
The summary of numerical experiments is shown in Table \ref{table 1}.
All the experiments have reached equilibrium. Equilibrium is determined through the time series analyses of domain-mean sea surface temperature and the balance between incoming shortwave radiation and outgoing longwave radiation at the top of the atmosphere (TOA). Experiments with higher stellar flux need a longer time to spin up. Each experiment runs for at least 1000 model days, and the data of the final 200 equilibrium model days are used for the analyses shown hereafter.

\section{Results} \label{results}
\subsection{Effects of Gravity on Planetary Climate under a Fixed Stellar Flux}\label{subsec: fixed}

\begin{figure}[t!]
\epsscale{1.1}
\plotone{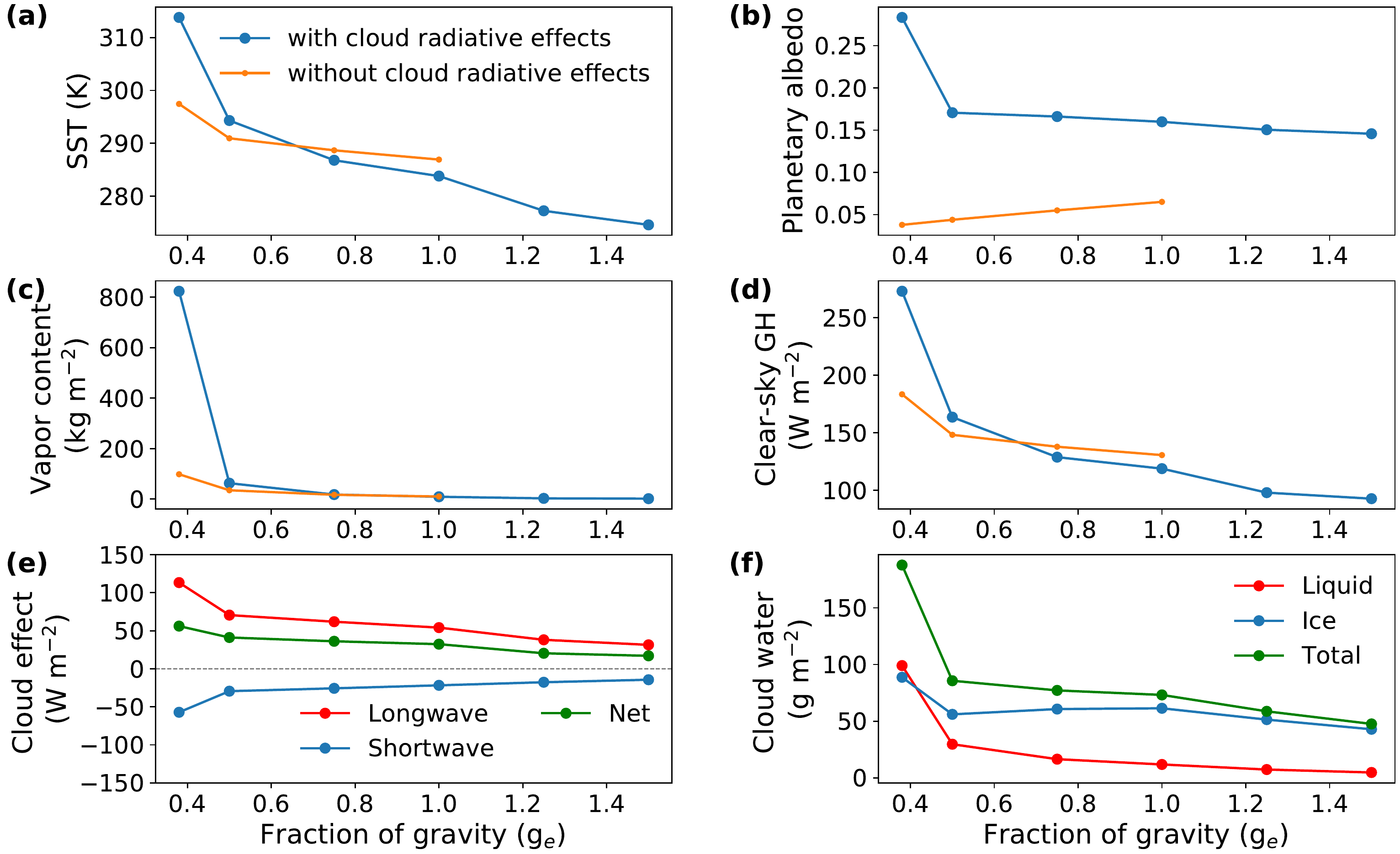}
\caption{Effects of varying planetary gravity on sea surface temperature (a), planetary albedo (b), column water vapor content (c), the strength of clear-sky greenhouse effect (d), cloud longwave (red line), shortwave (blue line), and net (green line) radiative effects at TOA (e), and column cloud liquid water (red line), cloud ice water (blue line), and total cloud water (green line) (f). The results are from experiments with a fixed stellar flux of 230 W\,m$^{-2}$, except the orange lines in panels (a)-(d) are results of experiments without cloud radiative effects and for a fixed stellar flux of 270 W\,m$^{-2}$. 
\label{effect_gravity_1}}
\end{figure}

\begin{figure}[t!]
\epsscale{0.9}
\plotone{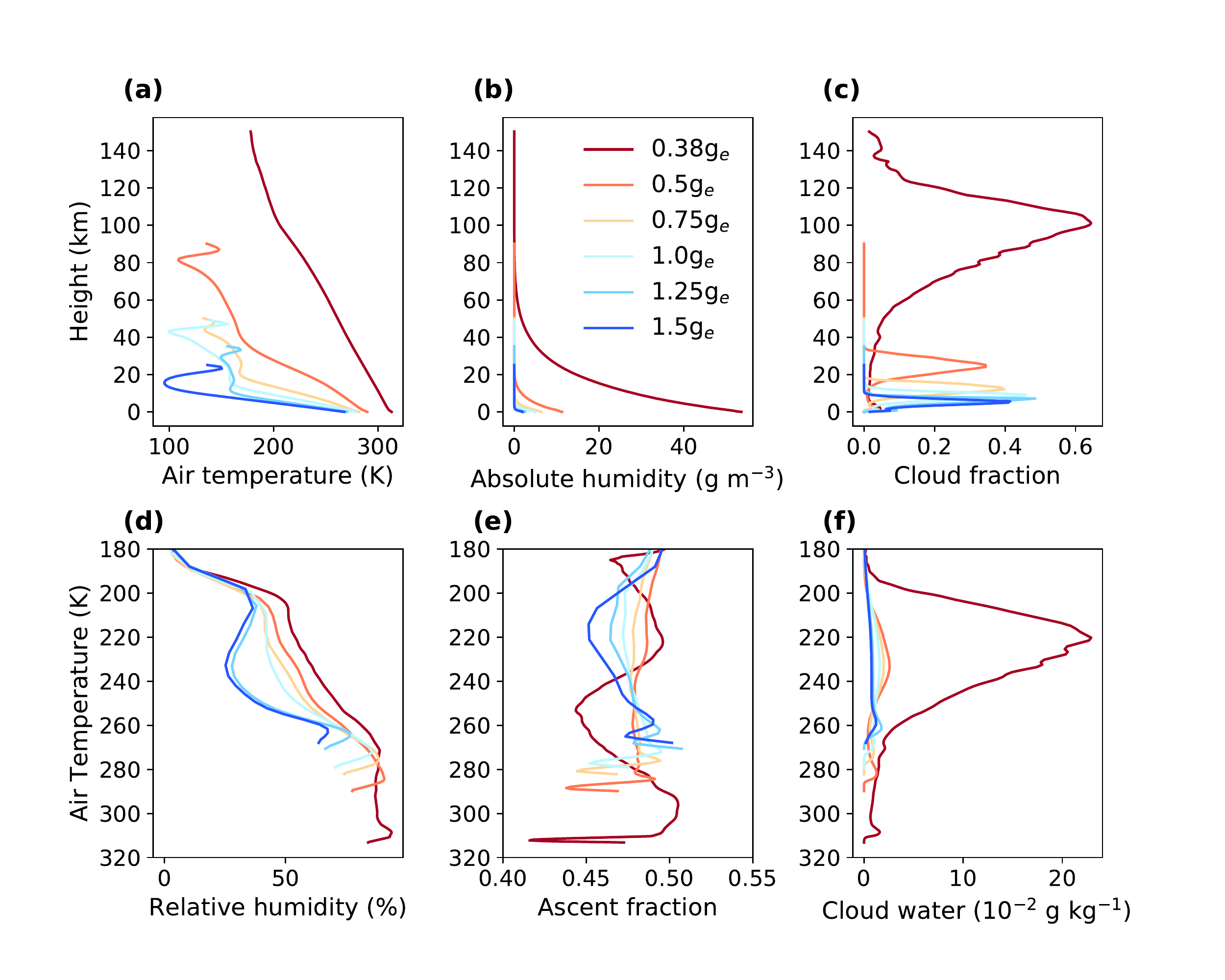}
\caption{Vertical profiles of domain- and time-mean air temperature (a), absolute humidity (b), cloud fraction (c), relative humidity (d), fraction of area occupied by updraft (e), and total cloud water (f) under a fixed stellar flux of 230 W\,m$^{-2}$. Since different experiments have different model-top heights, it is inconvenient to compare the variables in the altitude coordinate. Thus, to compare the variables more directly, for (d) to (f), we use air temperature as the vertical coordinate instead. 
\label{profile_1}}
\end{figure}

Our results show that sea surface temperature increases with decreasing gravity under small-domain RCE simulation (Figure \ref{effect_gravity_1}(a)). This is because water vapor (Figure \ref{effect_gravity_1}(c)) and cloud warming effect (Figure \ref{effect_gravity_1}(e)) increase with decreasing gravity. The total water vapor content in a column of atmosphere, W, can be given  by
\begin{equation}
    W 
    = \int_{0}^{\infty} \rho_{v} dz
    = \int_{0}^{\infty} \frac{\gamma e_{s}(T)}{R_{v}T} dz 
    = -\frac{1}{g}\frac{R}{R_{v}}\int_{p_s}^{0} \frac{\gamma e_{s}(T)}{p} dp
    = -\frac{1}{g}\frac{R}{R_{v}}\int_{0}^{\infty} \gamma e_{s}(T) d \ln (\frac{p}{p_{s}})
\label{water}
\end{equation}
where $\rho_{v}$ is the water vapor density in the atmosphere, $\gamma$ is the relative humidity, $e_{s}(T)$ is the saturated vapor pressure, $R_{v}$ and $R$ are the specific gas constants of water vapor and moist air, $T$ is the air temperature, $g$ is the planetary gravity, $p$ is the air pressure, and $p_{s}$ is the surface pressure. We use the ideal gas equation $\gamma e_{s}(T)=e=\rho_{v}R_{v}T$ to replace $\rho_{v}$, where $e$ is the water vapor pressure.
According to the Clausius-Clapeyron relationship, $e_{s}(T)$ is determined only by air temperature \citep{yau1996short}. To view the relation between the column water vapor content and the planetary gravity, we obtain the equation with an integral over air pressure $p$ using the hydrostatic equation $\frac{dp}{dz}=-\rho g$ and the ideal gas equation $p=\rho RT$, where $\rho$ is the density of moist air.
Then we can replace the integral over $p$ by an integral over $ln(p/p_{s})$ and get the final equation.
For given $e_{s}(T)$ and $\gamma$, the total water vapor content is approximately inversely proportional to the planetary gravity. 
Since water vapor is a potent greenhouse gas, the larger the water vapor content, the stronger the clear-sky greenhouse effect (Figure \ref{effect_gravity_1}(d)), and the warmer the climate. In addition, the warmer climate can further increase the water vapor content and the greenhouse effect, so the climate becomes even warmer, inducing a positive feedback. Therefore, a lower-gravity planet has a warmer climate.

To further verify the effect of water vapor content, we artificially switch off cloud radiative effects but still allow cloud formation (experiments without cloud radiative effects in Table \ref{table 1}). As shown with the orange lines in Figures \ref{effect_gravity_1}(a), (c), and (d), sea surface temperature, water vapor content, and clear-sky greenhouse effect increase monotonically with decreasing gravity, which confirms our discussions above. Without cloud radiative effects, planetary albedo decreases with decreasing gravity because more water vapor can absorb more incoming stellar flux in near-infrared wavelengths (the orange line in Figure \ref{effect_gravity_1}(b)).

The cloud warming effect increases with decreasing gravity (Figure \ref{effect_gravity_1}(e)), which further amplifies the warming. The domain in our simulations is too small for convective self-aggregation, since self-aggregation only occurs with domain lengths larger than 200-300 km \citep{https://doi.org/10.1002/grl.50204,muller2015favors}.
Convection evolves randomly and quasi-homogeneously over the domain in our simulations.
When gravity decreases, relative humidity increases (Figure \ref{profile_1}(d)). This is because in smaller gravity simulations the ascent area is larger in the upper troposphere (Figure \ref{profile_1}(e)). Higher relative humidity and a larger fraction of ascent area promote cloud formation, which leads to a larger cloud fraction (Figure \ref{profile_1}(c)) and more cloud condensate in smaller gravity simulations (Figures \ref{effect_gravity_1}(f) and \ref{profile_1}(f)). The increasing cloud condensate increases cloud optical depth, resulting in a larger planetary albedo (Figure \ref{effect_gravity_1}(b)) and stronger cloud shortwave and longwave radiative effects (Figure \ref{effect_gravity_1}(e)). The change of the cloud longwave radiative effect dominates that of the cloud shortwave radiative effect, so the cloud net radiative effect increases with decreasing gravity (Figure \ref{effect_gravity_1}(e)). 
Cloud radiative effects depend on various factors, including cloud fraction, cloud water, cloud temperature, lapse rate, and cloud particle size \citep{hartmann1994global}. However, none of them alone can provide a clear tendency to explain why the cloud longwave radiative effect changes more than the cloud shortwave effect in our simulations, so there may be a complex combined effect here.

It is worth noting that clouds have a net warming effect in our experiments, however, they have a global-mean net cooling effect in Earth's observations and global-scale simulations \citep[e.g.,][]{ramanathan1989cloud,allan2011combining,https://doi.org/10.1029/2022GL100152}. Two reasons may contribute to this difference. First, in small domain RCE, there is no large-scale subsidence or cloud self-aggregation \citep{AnEnergyBalanceAnalysisofDeepConvectiveSelfAggregationaboveUniformSST}, so more high-level clouds form and these clouds perform strong warming effect. As seen in Figures \ref{profile_1}(c) and \ref{effect_gravity_1}(f), the cloud fraction of high-level clouds is much larger than the low-level clouds, and there is more cloud ice water than cloud liquid water in our simulations. Second, compared with other studies \citep[e.g.,][]{wing2020clouds,seeley2021episodic}, the stellar flux is lower in our experiments (more discussions are shown in Section \ref{conclusions}), so with the same cloud albedo, clouds reflect less shortwave radiation to space, resulting in weaker cloud shortwave radiative effects. Due to the limitation of computation resources, we do not perform experiments in a larger domain. Future works including large-scale air motions are expected to verify our results.

\begin{figure}[t!]
\epsscale{1.05}
\plotone{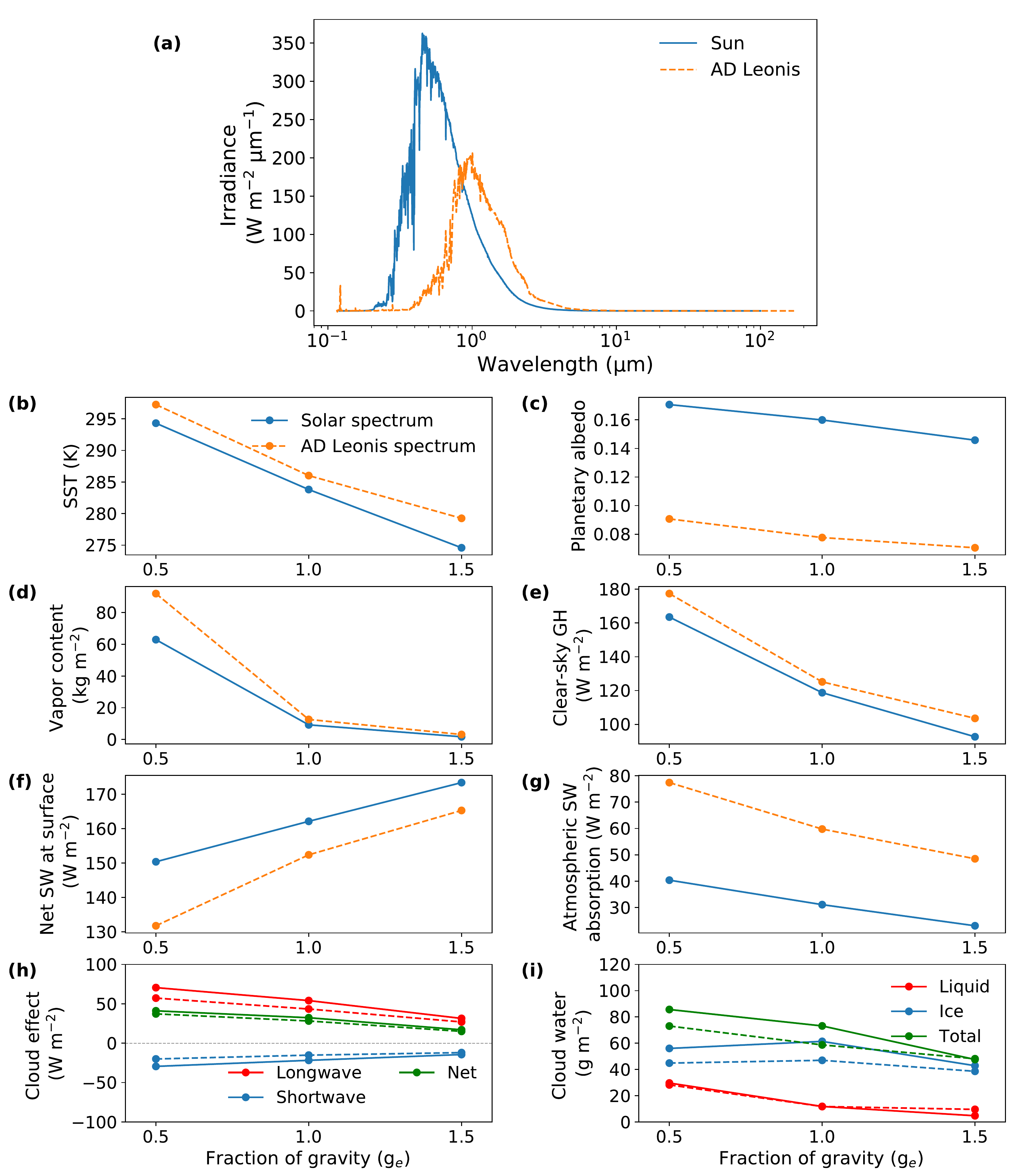}
\caption{Stellar spectra for the Sun (blue solid line) and AD Leonis (orange dash line), with a fixed total stellar flux of 230 W\,m$^{-2}$ (a).
Effects of varying planetary gravity with the solar spectrum (blue solid lines) and the AD Leonis spectrum (orange dash lines) on sea surface temperature (b), planetary albedo (c), column water vapor content (d), the strength of clear-sky greenhouse effect (e), the net shortwave flux at the surface (f), and atmospheric shortwave absorption (g). 
Effects of varying planetary gravity on cloud longwave (red line), shortwave (blue line), and net (green line) radiative effects at TOA (h), and column cloud liquid water (red line), cloud ice water (blue line), and total cloud water (green line) (i) for the experiments with the AD Leonis spectrum (dash lines) and the solar spectrum (solid lines). The results with the solar spectrum are the same as those shown in Figure \ref{effect_gravity_1}, and the stellar flux is 230 W\,m$^{-2}$ for all the experiments.}
\label{effect_gravity_1_Mstar}
\end{figure}

It is now known that the majority of stars in the galaxy are M stars, and exoplanets may be common in the habitable zone of M stars \citep[e.g.,][]{Winters_2014}. To extend our study to more terrestrial exoplanets, we perform an additional group of experiments with an M star spectrum. The M star that we choose is AD Leonis, a main sequence star with a spectral classification of M3.5V. The effective temperature of AD Leonis is about 3400 K, much lower than the Sun, which has an effective temperature of about 5770 K. The spectra for the Sun and AD Leonis are shown in Figure \ref{effect_gravity_1_Mstar}(a). One can see that because the Sun has a higher effective temperature, its emission peak moves to shorter wavelengths than AD Leonis. The solar irradiance peaks at visible wavelengths, whereas AD Leonis peaks at near-infrared wavelengths.

The effects of gravity with the AD Leonis spectrum are shown in Figures \ref{effect_gravity_1_Mstar}(b)-(i). The results show that sea surface temperature increases with decreasing gravity (Figure \ref{effect_gravity_1_Mstar}(b)). This is because water vapor content, clear-sky greenhouse effect, and cloud net radiative effect increase with decreasing gravity (Figures \ref{effect_gravity_1_Mstar}(d), (e), and (h)), same as the discussion with the solar spectrum above. This confirms that our conclusions are not sensitive to stellar spectrum.

Comparing the results of the solar spectrum and the AD Leonis spectrum, one can see that sea surface temperature is larger with the AD Leonis spectrum than with the solar spectrum under the same gravity value. This is because the atmosphere, especially water vapor and carbon dioxide can absorb more incoming stellar irradiation in the near-infrared wavelengths. This leads to increased atmospheric shortwave absorption (Figure \ref{effect_gravity_1_Mstar}(g)), decreased net surface shortwave flux (Figure \ref{effect_gravity_1_Mstar}(f)), and decreased planetary albedo (Figure \ref{effect_gravity_1_Mstar}(c)). As the atmosphere absorbs more stellar irradiation, air temperature increases (figure not shown), and so does atmospheric stability, which reduces cloud formation, resulting in less cloud cover (figure not shown) and less cloud water path (Figure \ref{effect_gravity_1_Mstar}(i)). The reduction in cloud cover reduces cloud net radiative effect by 13.29, 10.71, and 4.47 W\,m$^{-2}$ for the experiments of 0.5g$_{e}$, 1.0g$_{e}$, and 1.5g$_{e}$, respectively (Figure \ref{effect_gravity_1_Mstar}(h)). Planetary albedo decreases by 0.080, 0.082, and 0.075 for the experiments of 0.5g$_{e}$, 1.0g$_{e}$, and 1.5g$_{e}$, which increases the net incoming shortwave radiation by 18.40, 18.93, and 17.32 W\,m$^{-2}$. Thus, the increase in net incoming shortwave radiation overweighs the reduction in cloud net radiative effect, resulting in a warmer climate with the AD Leonis spectrum than the solar spectrum.

\subsection{Effects of Gravity on Planetary Climate with Increasing Stellar Flux}\label{subsec: increasing}

\begin{figure}[t!]
\epsscale{1.15}
\plotone{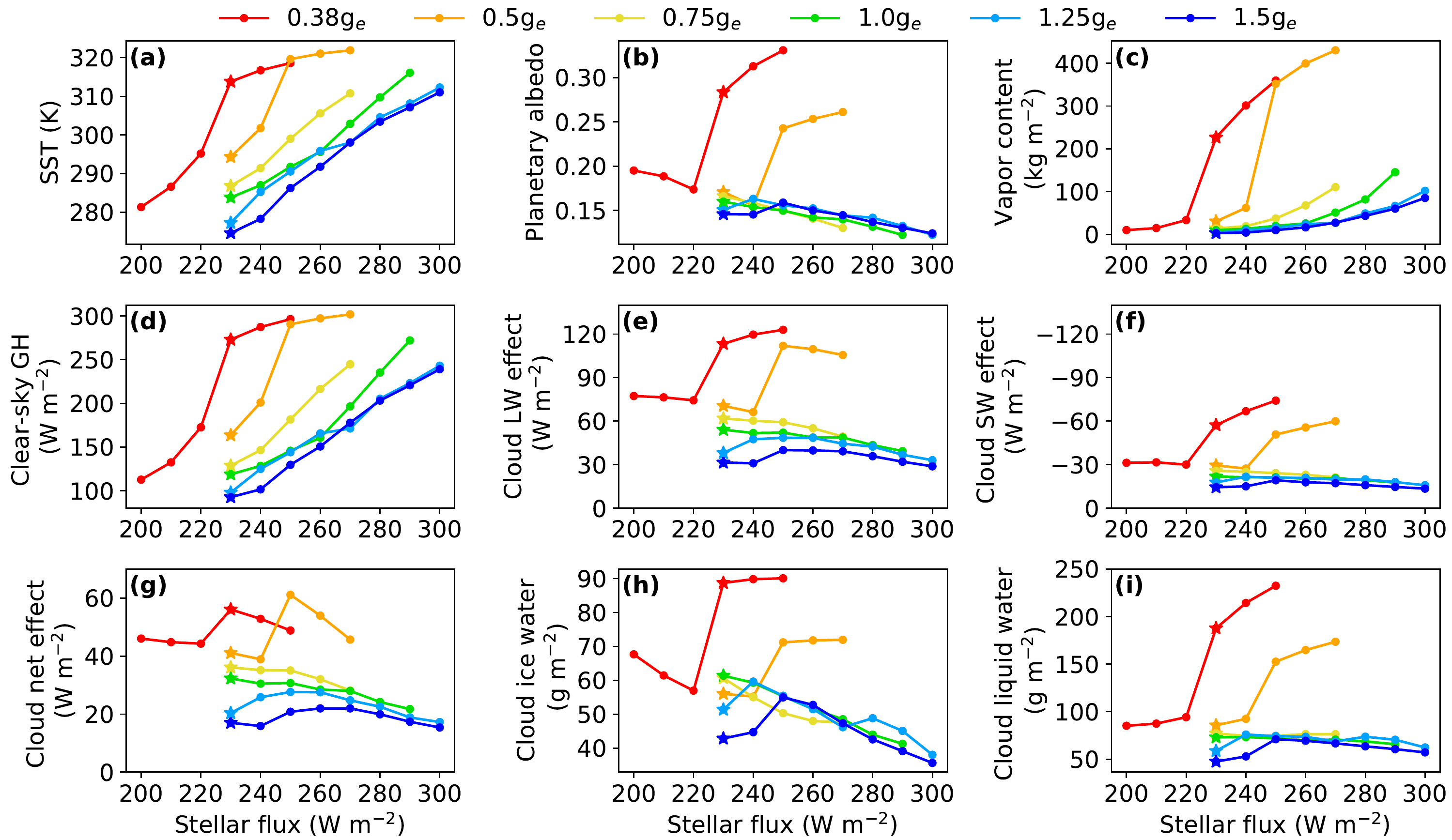}
\caption{Effects of gravity and stellar flux on sea surface temperature (a), planetary albedo (b), column water vapor content (c), the strength of clear-sky greenhouse effect (d), cloud longwave radiative effect (e), cloud shortwave radiative effect (f), cloud net radiative effect (g), column cloud ice water path (h), and column cloud liquid water path (i). The control value of 230 W\,m$^{-2}$ in Section \ref{subsec: fixed} is marked by a star. For the cases of 0.5g$_e$, 0.75g$_e$, 1.0g$_e$, 1.25g$_e$, and 1.5g$_e$ we increase the stellar flux from 230 W\,m$^{-2}$ with an increment of 10-W\,m$^{-2}$ until the model crashes under high surface temperatures. For the cases of 0.38g$_e$, convection has entered the oscillatory state with 230 W\,m$^{-2}$, so we start to increase the stellar flux from 200 W\,m$^{-2}$ to simulate the quasi-steady state.}
\label{effect_gravity_2}
\end{figure}

\begin{figure}[t!]
\epsscale{1.15}
\plotone{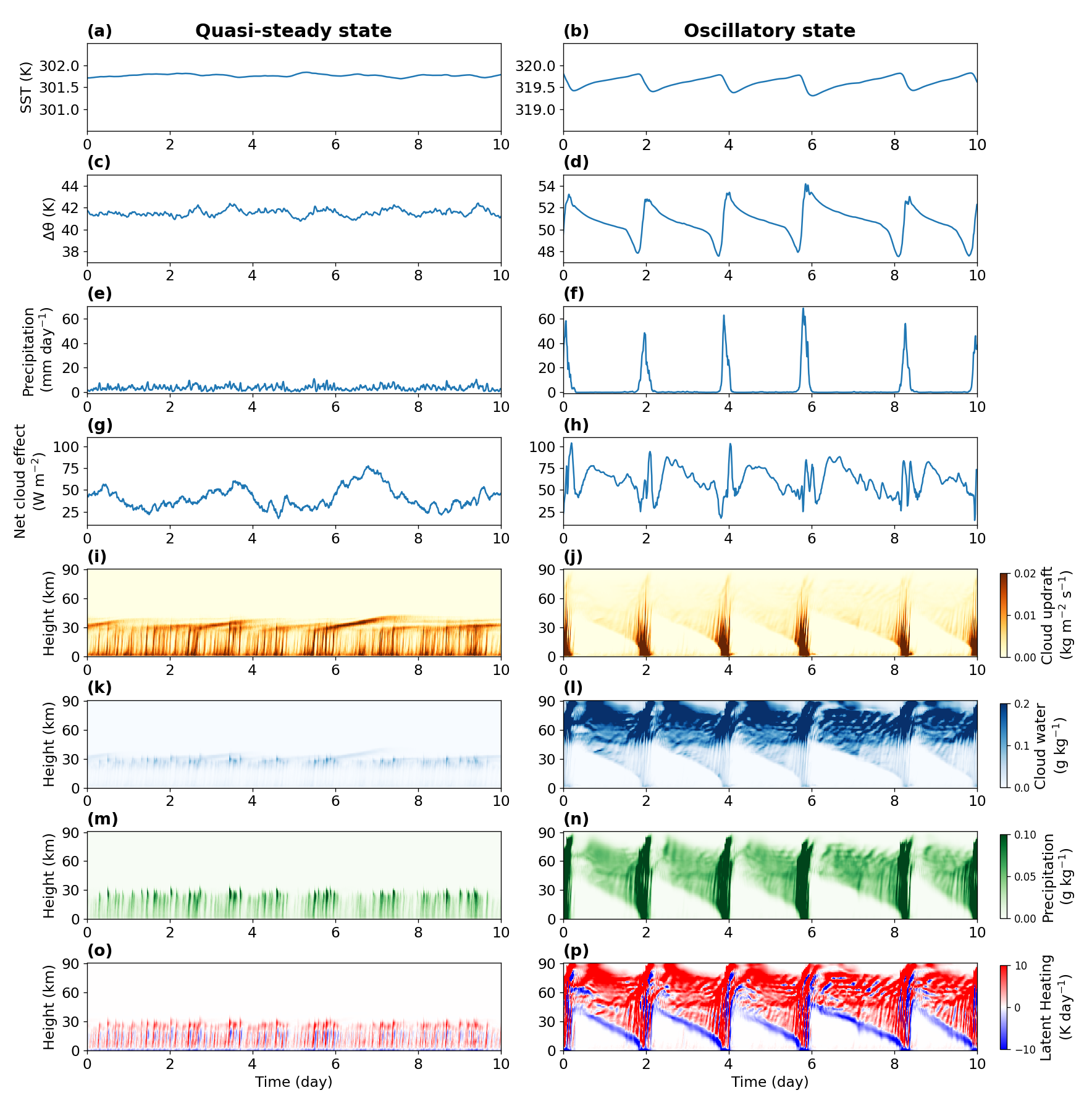}
\caption{Time series of domain-mean sea surface temperature (a-b), the potential temperature difference between the lower troposphere (10 to 15 km) and the surface (lower troposphere minus surface, c-d), surface precipitation (e-f), and net cloud radiative effect at TOA (g-h). Time-versus-height plots of domain-mean updraft cloud mass flux (i-j), cloud water (k-l), precipitation (m-n), and latent heating rate (including both water vapor condensation and re-evaporation of precipitating droplets, o-p). The panels on the left are the behaviors of a quasi-steady state (0.5g$_e$ with 240\, W\,m$^{-2}$), and the panels on the right are the behaviors of an oscillatory state (0.5g$_e$ with 250 W\,m$^{-2}$). 
\label{time_series}}
 \end{figure}

\begin{figure}[t!]
\epsscale{1.1}
\plotone{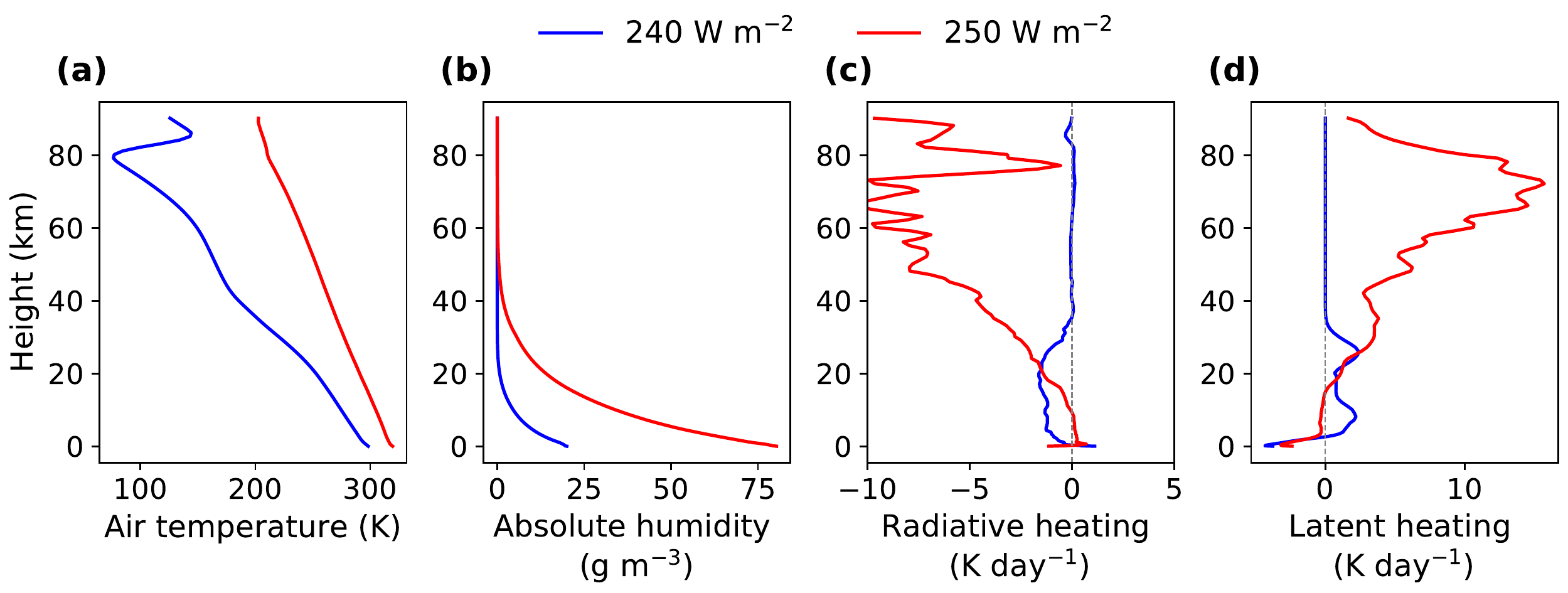}
\epsscale{0.4}
\caption{Vertical profiles of domain- and time-mean air temperature (a), absolute humidity (b), radiative heating (c), and latent heating (d). The blue lines are the results of the case of 0.5g$_e$ with 240 W\,m$^{-2}$, an example of the quasi-steady state, and the red lines are the results of the case of 0.5g$_e$ with 250 W\,m$^{-2}$, an example of the oscillatory state.
\label{profile_0.5g}}
\end{figure}

\begin{figure}[t]
\epsscale{1.1}
\plotone{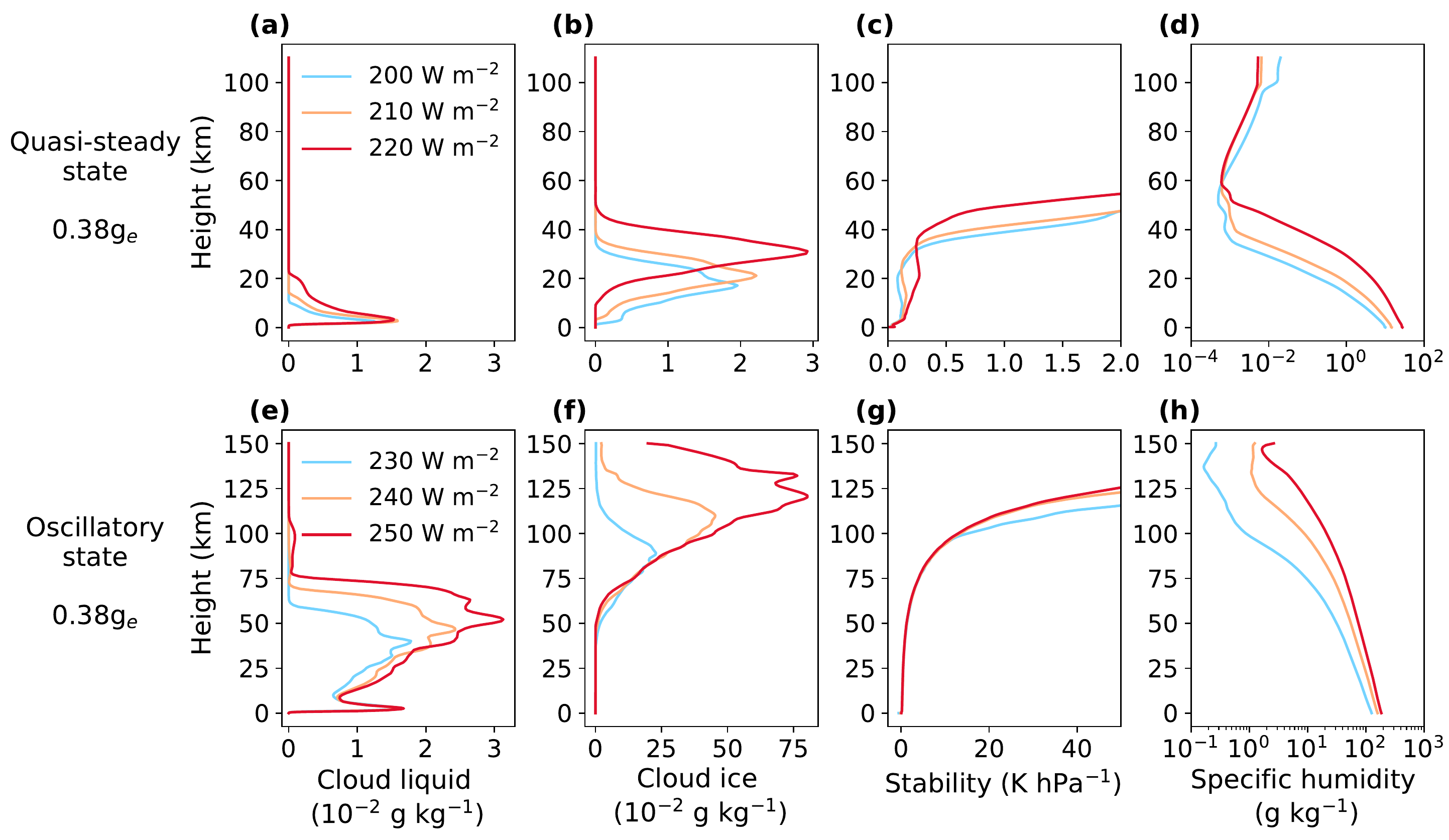}
\epsscale{-1}
\caption{Vertical profiles of domain- and time-mean cloud liquid water (a and e), cloud ice water (b and f), stability (c and g), and specific humidity (d and h) for the cases of 0.38g$_e$. The stability is calculated by $-\frac{T}{\theta} \frac{\partial \theta}{\partial p}$, where $T$ is the air temperature, $\theta$ is the potential temperature, and $p$ is the air pressure \citep{emanuel1994large,doi:10.1073/pnas.1601472113}. The upper row presents the results in the quasi-steady state, and the lower row presents the results in the oscillatory state.}
\label{profile_posttransition}
\end{figure}

As seen in Figure \ref{effect_gravity_2}(a), under a given gravity value, sea surface temperature increases with stellar flux, and planets of different values of gravity exhibit different increasing rates. For higher-gravity planets, 0.75g$_e$, 1.0g$_e$, 1.25g$_e$, and 1.5g$_e$, the increasing rate is nearly a constant at least in the range of the stellar flux examined in our simulations. For lower-gravity planets, 0.38g$_e$ and 0.5g$_e$, the increasing rate is first near a constant, then increases, and then decreases.

As seen in Figures \ref{effect_gravity_2}(c) and (d), for the cases of 0.75g$_e$, 1.0g$_e$, 1.25g$_e$, and 1.5g$_e$, column water vapor content and clear-sky greenhouse effect increase with stellar flux, due to the positive water vapor feedback. 
Cloud net radiative effect has an overall decreasing tendency in all the experiments of 0.75g$_e$ and 1.0g$_e$ and in the experiments of 1.25g$_e$ and 1.5g$_e$ when the stellar flux is equal to or above 260 W\,m$^{-2}$ (Figure \ref{effect_gravity_2}(g)). This is because as the climate gets warmer and wetter, the environmental lapse rate decreases, indicating a more stable atmosphere; the enhanced static stability weakens deep convection and decreases high cloud formation \citep{doi:10.1073/pnas.1601472113}. 
With enough water vapor source, low cloud formation is less affected by the enhanced static stability, so cloud liquid water (Figure \ref{effect_gravity_2}(i)) decreases less than cloud ice water (Figure \ref{effect_gravity_2}(h)). Thus, the decrease in the cloud longwave radiative effect dominates, and the cloud net radiative effect decreases with stellar flux (Figure \ref{effect_gravity_2}(g)).
Overall, planetary albedo decreases with increasing stellar flux (Figure \ref{effect_gravity_2}(b)) because more water vapor can absorb more incoming stellar flux in near-infrared wavelengths, and meanwhile, the cloud liquid water and cloud ice water contents decrease.

For the cases of 0.38g$_e$ and 0.5g$_e$, the increasing rates of sea surface temperature change because convection shifts from a quasi-steady state (left panels of Figure \ref{time_series}) to an oscillatory state (right panels of Figure \ref{time_series}) as the stellar flux increases.
In the quasi-steady state, the thermal and water variables display random fluctuations. In the oscillatory state, there is a two-day convective cycle, comprised of a dry phase and a wet phase (right panels of Figure \ref{time_series}). As seen in Figures \ref{time_series}(f), (j), and (l), in the wet phase, surface precipitation occurs as an outburst, and convection can extend from the surface to the upper troposphere; in the dry phase, there is no surface precipitation or near-surface convection. Compared with the quasi-steady state, in the oscillatory state, both water vapor content (Figure \ref{profile_0.5g}(b)) and cloud water (Figures \ref{time_series}(k) and (l)) increase dramatically, and clouds can rise to higher altitudes (Figure \ref{time_series}(l)), resulting in a larger clear-sky greenhouse effect (red and orange lines in Figure \ref{effect_gravity_2}(d)) and a stronger cloud warming effect (red and orange lines Figures in \ref{effect_gravity_2}(g) and \ref{time_series}(h)). The combination of these two effects causes rapid warming (red and orange lines in Figure \ref{effect_gravity_2}(a)). 

As discussed in \citet{seeley2021episodic}, the dry phase of the oscillatory state may be caused by a positive radiative heating rate in the lower troposphere. When water vapor content is very large in the lower troposphere, the thermal infrared absorption window regions close off, so the lower atmosphere can not be cooled by emitting thermal infrared \citep{pierrehumbert2010principles,wordsworth2013water}. Meanwhile, water vapor can still absorb near-infrared radiation of the incoming stellar flux. This leads to an overall positive radiative heating rate in the lower troposphere (red line in Figure \ref{profile_0.5g}(c)). Thus, the potential temperature difference between the lower troposphere and surface air increases (Figure \ref{time_series}(d)), and the environmental lapse rate decreases (Figure \ref{profile_0.5g}(a)), which enhances static stability in the lower troposphere and suppresses surface-based convection (Figures \ref{time_series}(j) and (l)). On the contrary, radiative heating in the lower troposphere enhances instability in the upper troposphere, so there is still elevated convection, whose precipitation evaporates before reaching the surface (Figures \ref{time_series}(j), (l), and (n)). Elevated convection moves lower due to the evaporative cooling of precipitation (Figure \ref{time_series}(p)).
When the elevated convection moves low enough for the evaporation of precipitation to cool the lower troposphere (Figure \ref{time_series}(p)), the potential temperature difference between the lower troposphere and the surface decreases (Figure \ref{time_series}(d)), and the suppression is eliminated to allow surface-based convection to evaluate to higher altitudes and produce heavy rainfall (Figures \ref{time_series}(j), (l), and (n)). The evaporation of the heavy rainfall causes a time-mean net latent cooling in the lower troposphere (Figure \ref{profile_0.5g}(d)). Note that the period of the convection cycle varies between different simulations, for example, the period is about two days in the case of 0.5g$_e$ with 250 W\,m$^{-2}$, and is about one and a half days in the case of 0.38g$_e$ with 250 W\,m$^{-2}$ (figure not shown). Further works are expected to investigate the mechanisms that determine the period of the convection cycle.
The simulations in \citet{seeley2021episodic} are under Earth's gravity value, while our results suggest that the mechanism of the oscillatory state should be universal on terrestrial exoplanets despite their different values of planetary gravity. 

A lower-gravity planet requires a relatively cooler climate to shift to the oscillatory state.
For the cases of 0.38g$_e$ and 0.5g$_e$, the threshold temperatures are below 315 K and 320 K, respectively. For the Earth's gravity value, the transition requires an extremely warm hothouse climate, with the surface temperature reaching about 325 K \citep{seeley2021episodic}. This is likely because water vapor content increases with decreasing gravity in the whole column as well as in each layer according to Equation (\ref{water}), so a lower-gravity planet requires a cooler climate to have a positive radiative heating rate in the lower troposphere. Note that due to the model limitation and the computation resource limitation, the surface temperature can not be high enough for larger gravity simulations to shift to the oscillatory state in our simulations.

\begin{figure}[t!]
\epsscale{1.15}
\plotone{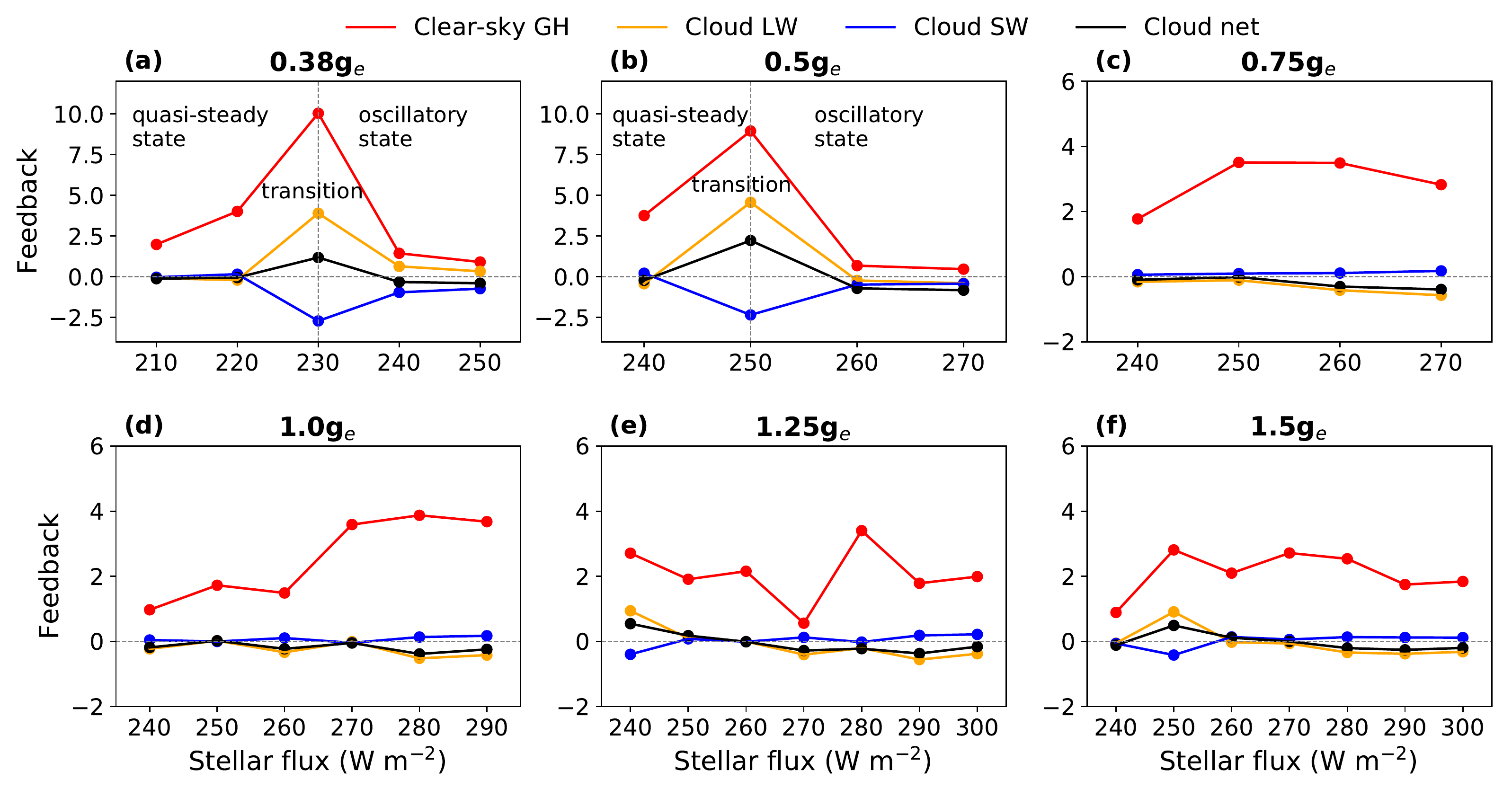}
\epsscale{-1}
\caption{Clear-sky greenhouse effect feedback (red lines), cloud longwave radiative feedback (orange lines), cloud shortwave radiative feedback (blue lines), and net cloud radiative feedback (black lines) as a function of stellar flux for the cases of 0.38g$_e$ (a), 0.5g$_e$ (b), 0.75g$_e$ (c), 1.0g$_e$ (d), 1.25g$_e$ (e), and 1.5g$_e$ (f). The feedback parameter is defined as the change in clear-sky greenhouse effect or cloud radiative effects to the increase of stellar flux (W\,m$^{-2}$ per W\,m$^{-2}$).
\label{feedback}}
\end{figure}

Once the system enters the oscillatory state, the oscillatory regime persists as stellar flux increases (figure not shown). Sea surface temperature grows slowly compared to the quasi-steady state (Figure \ref{effect_gravity_2}(a)). This is because the cloud liquid water increases more and cloud shortwave radiative effect strengthens faster with stellar flux in the oscillatory state (red and orange lines in Figures \ref{effect_gravity_2}(i) and (f)). In the oscillatory state, the atmospheric stability in the low- and mid-troposphere changes little with stellar flux (Figure \ref{profile_posttransition}(g)); while in the quasi-steady state, the atmospheric stability increases with stellar flux (Figure \ref{profile_posttransition}(c)), which discourages cloud formation. Thus, cloud liquid water increases more with stellar flux in the oscillatory state (Figures \ref{profile_posttransition}(a) and (e)). More cloud liquid water leads to a stronger cloud shortwave radiative effect, so cloud shortwave radiative effect also strengthens more with stellar flux in the oscillatory state.
In the oscillatory state, cloud ice water moves upward as stellar flux increases (Figure \ref{profile_posttransition}(f)). However, since the atmosphere is thin in high altitudes, the total cloud ice water path remains nearly unchanged (Figure \ref{effect_gravity_2}(h)). This results in a slow increase in cloud longwave radiative effect in the cases of 0.38g$_e$ and a decrease in the cases of 0.5g$_e$ as stellar flux increases (Figure \ref{effect_gravity_2}(e)).
The rapid strengthening of cloud shortwave radiative effect leads to a faster reduction in cloud net radiative effect as stellar flux increases in the oscillatory state, stabilizing the climate.

To estimate the climate stability with increasing stellar flux, we calculate the feedbacks of the clear-sky greenhouse effect and cloud radiative effects to stellar flux change (Figure \ref{feedback}). The feedback is calculated by
\begin{equation}
    F = \frac{\Delta E}{\Delta S}
\end{equation}
where $\Delta E$ is the incremental change in cloud radiative effects or clear-sky greenhouse effect in W\,m$^{-2}$ and $\Delta S$ is the increment of stellar flux in W\,m$^{-2}$.
As shown in Figures \ref{feedback}(c)-(f), for the cases of 0.75g$_e$, 1.0g$_e$, 1.25g$_e$, and 1.5g$_e$, the clear-sky radiative effect feedback is always positive; the cloud shortwave radiative feedback is positive, the cloud net radiative feedback and the cloud longwave radiative feedback are negative except for the cases in 1.25g$_e$ and 1.5g$_e$ below 260 W\,m$^{-2}$. This means that water vapor causes a positive feedback to destabilize the climate and clouds have an overall negative feedback to stabilize the climate under the quasi-steady state in small-domain simulations.
For the cases of 0.38g$_e$ and 0.5g$_e$, during the transition from the quasi-steady state to the oscillatory state, the net cloud radiative feedback to stellar flux changes from near-zero to a positive value, and the clear-sky greenhouse feedback increases substantially (Figures \ref{feedback}(a) and (b)). These two effects combined to cause rapid warming. 
When the stellar flux increases under the oscillatory state, both the clear-sky greenhouse feedback and the cloud net radiative feedback decrease, so sea surface temperature grows slowly. 

In the oscillatory state, the climate has entered the moist greenhouse state \citep{kasting1993habitable}, in which the specific humidity reaches or exceeds 3 g\,kg$^{-1}$ in the stratosphere (Figure \ref{profile_posttransition}(h)). This sufficient wet stratosphere may accelerate the water vapor loss, hastening the loss of the planetary ocean.

\subsection{Sensitivity Tests}\label{sec: sensitivity}

\begin{figure}[t]
\epsscale{0.95}
\plotone{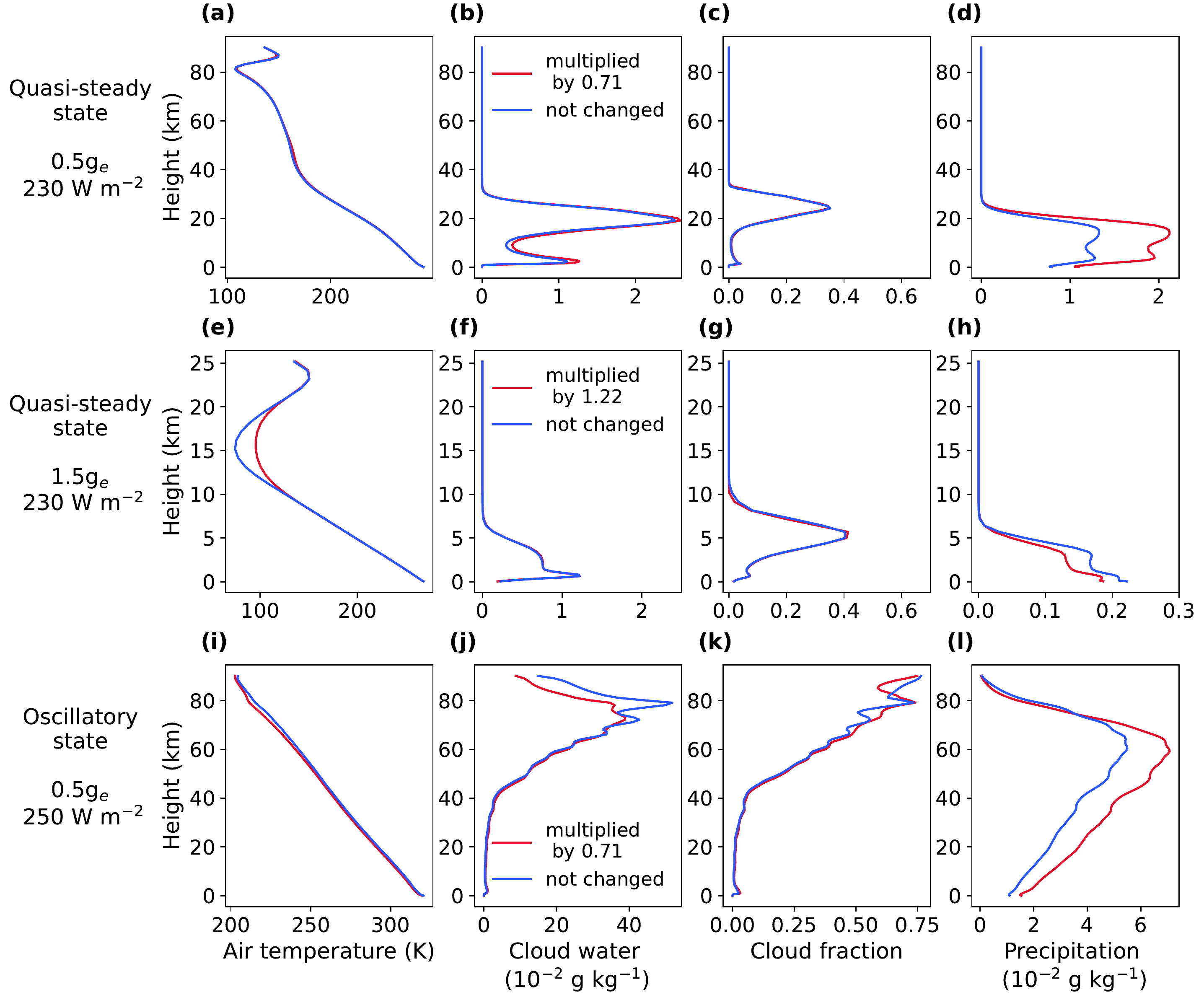}
\caption{The results of the sensitivity tests to the fall speeds of precipitation droplets. The upper row is the results of the case of 0.5g$_e$ with 230 W\,m$^{-2}$, in the quasi-steady state; the middle row is the results of the case of 1.5g$_e$ with 230 W\,m$^{-2}$, in the quasi-steady state; and the third row is the results of the case of 0.5g$_e$ with 250 W\,m$^{-2}$, in the oscillatory state. From left to right in all rows: vertical profiles of domain- and time-mean air temperature, cloud water, cloud fraction, and precipitation. The red lines are the results of the control runs and the blue lines are the results of the sensitivity tests.}
\label{fallspeed}
\end{figure}
\begin{figure}
\epsscale{0.95}
\plotone{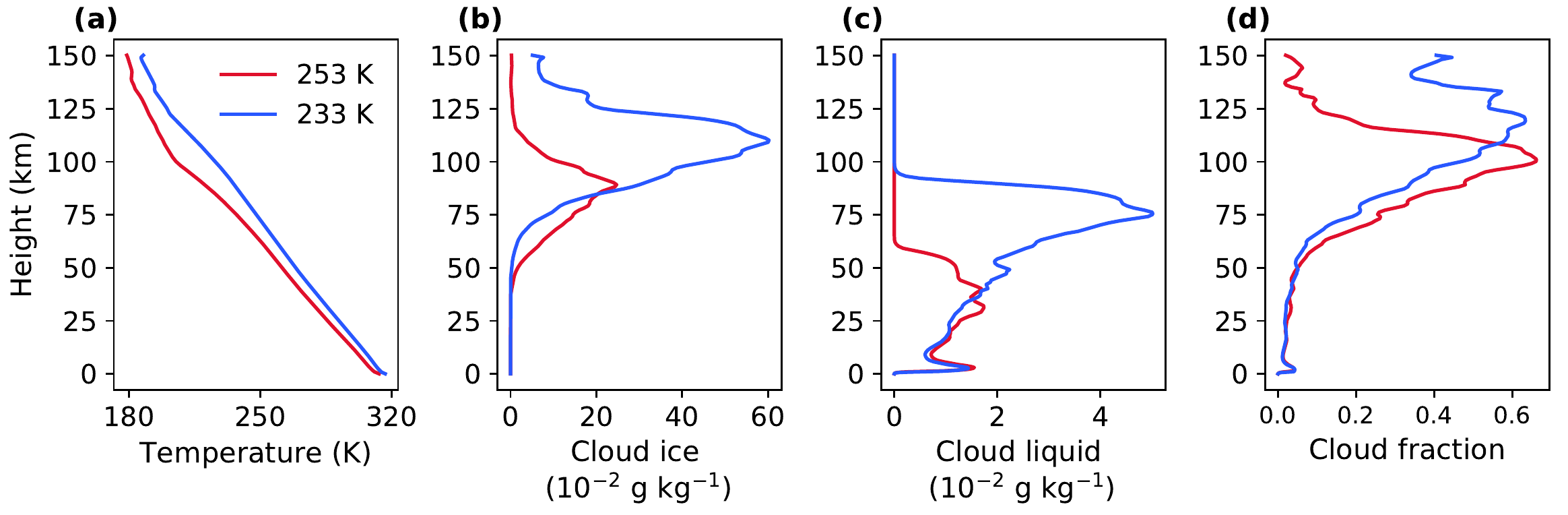}
\caption{The results of the sensitivity tests to the threshold temperature of cloud ice water and cloud liquid water, 0.38g$_e$ with 230 W\,m$^{-2}$. Vertical profiles of domain- and time-mean air temperature (a), cloud ice water (b), cloud liquid water (c), and cloud fraction (d). The red lines are for the control run, with a threshold temperature of 253 K, and the blue lines are for the sensitivity test, with a threshold temperature of 233 K.}
\label{icecloud}
\end{figure}

As stated in Section \ref{methods}, the fall speeds of precipitation droplets should change with varying gravity, and we have changed the constant fall speeds in SAM according to gravity in the simulations. To estimate the sensitivity to fall speeds, we add three experiments, without changing the default fall speeds in SAM, two in the quasi-steady state: 0.5g$_e$ with 230 W\,m$^{-2}$ and 1.5g$_e$ with 230 W\,m$^{-2}$, and one in the oscillatory state: 0.5g$_e$ with 250 W\,m$^{-2}$. The results are shown in Figure \ref{fallspeed}.
In the two experiments of the quasi-steady state, the fall speeds of precipitation droplets have minor effects on air temperature, cloud water, and cloud fraction profiles, but it changes the precipitation profiles in the lower troposphere (Figures \ref{fallspeed}(a) to (h)). As the fall speeds decrease, precipitation increases in each layer (Figures \ref{fallspeed}(d) and (h)). This is because, with smaller fall speeds, precipitation particles do not fall much and therefore stay more suspended in the air. In the experiments of the oscillatory state, fall speeds also have minor effects on air temperature, cloud water, and cloud fraction profiles (Figures \ref{fallspeed}(i) to (k)), but as fall speeds decrease, precipitation increases (Figure \ref{fallspeed}(l)). Convection still enters the oscillatory state as fall speeds change (figure not shown).
Our results indicate that the fall speeds of precipitation droplets have moderate effects on precipitation, but minor effects on cloud formation. The transition to the oscillatory state is not sensitive to fall speeds.

The partitioning of diagnosed cloud liquid water and cloud ice water is based on air temperature on every time step, and the default threshold in SAM is 253 K (below 253 K, cloud water is only comprised of cloud ice water), which is 20 K higher than most GCMs \citep{collins2004description,collins2006formulation}. To demonstrate the effects of this threshold temperature, we do an additional experiment of 0.38g$_e$ with 230 W\,m$^{-2}$, in which we artificially decrease the threshold temperature to 233 K. 
As the threshold temperature decreases, the partition between cloud ice water and cloud liquid water changes, with cloud liquid water occurring at higher altitudes (Figure \ref{icecloud}(c)). The column cloud liquid water path increases from 99 to 126 g m$^{-2}$, and cloud ice water path decreases from 88 to 71 g m$^{-2}$. The increase of cloud liquid water reduces net upward longwave radiation by about 20 W\,m$^{-2}$, while the change of cloud ice water has a small impact on net upward longwave radiation. Although the increase of cloud liquid water also increases planetary albedo and reduces net downward shortwave radiation (figure not shown), its warming effect dominates the results, leading to increased sea surface temperature and air temperature (Figure \ref{icecloud}(a)). Clouds can extend to high altitudes under a lower threshold temperature (Figure \ref{icecloud}(d)) and convection still enters the oscillatory state, implying that our results are robust.

\subsection{Transmission Spectra}\label{sec: transmission}

Transmission spectroscopy is effective in characterizing atmospheric components by estimating how much stellar irradiation is absorbed or scattered by the atmosphere on the exoplanets in the primary transit. The transmission spectra of hot giant planets and warm Neptune-size planets have been studied by the Hubble Space Telescope (HST) and the Spitzer space telescope \citep[e.g.,][]{Kreidberg_2014,knutson2014featureless,fraine2014water,Kreidberg_2015}. With the James Webb Space Telescope (JWST), scientists can get transmission spectra with high resolution, and the atmospheric components of terrestrial planets around small stars might be detectable \citep{Greene_2016,Morley_2017,Lustig_Yaeger_2019,Lacy_2020}.

\begin{figure}[ht!]
    \epsscale{1.1}
    \plotone{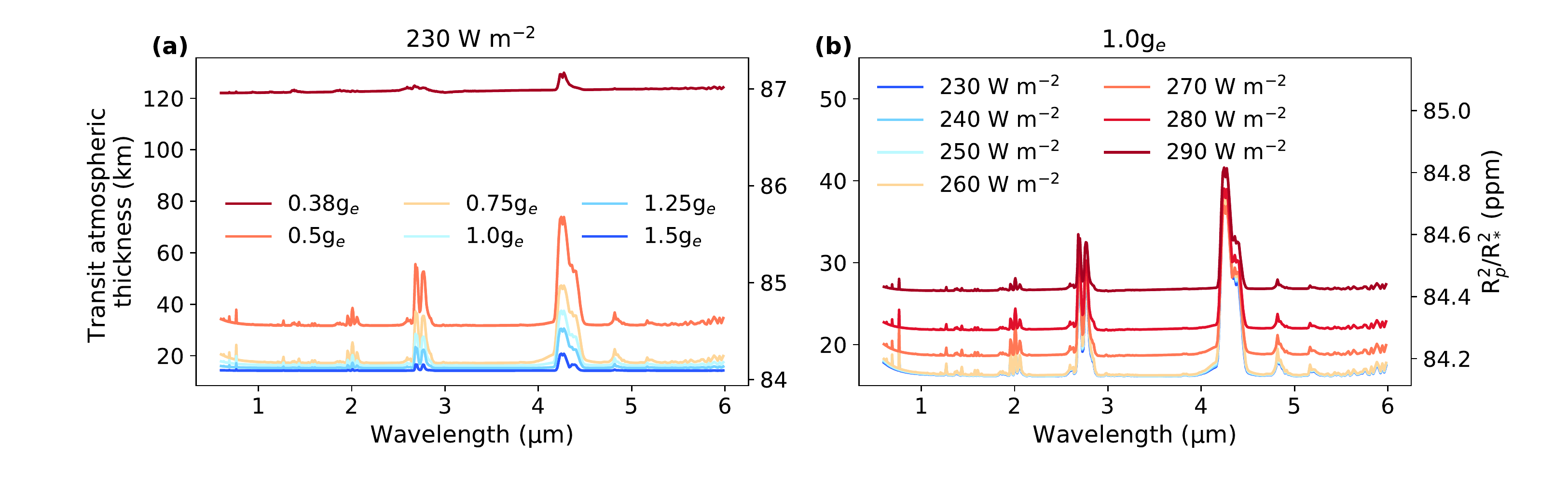}
    \epsscale{0.4}
    \caption{Domain- and time-mean transmission spectra for different gravity values with a fixed stellar flux of 230 W\,m$^{-2}$ (a) and increasing stellar flux with a fixed gravity value of 1.0g$_e$ (b).}
    \label{transmission_all}
\end{figure}
    
\begin{figure}
    \epsscale{1.15}
    \plotone{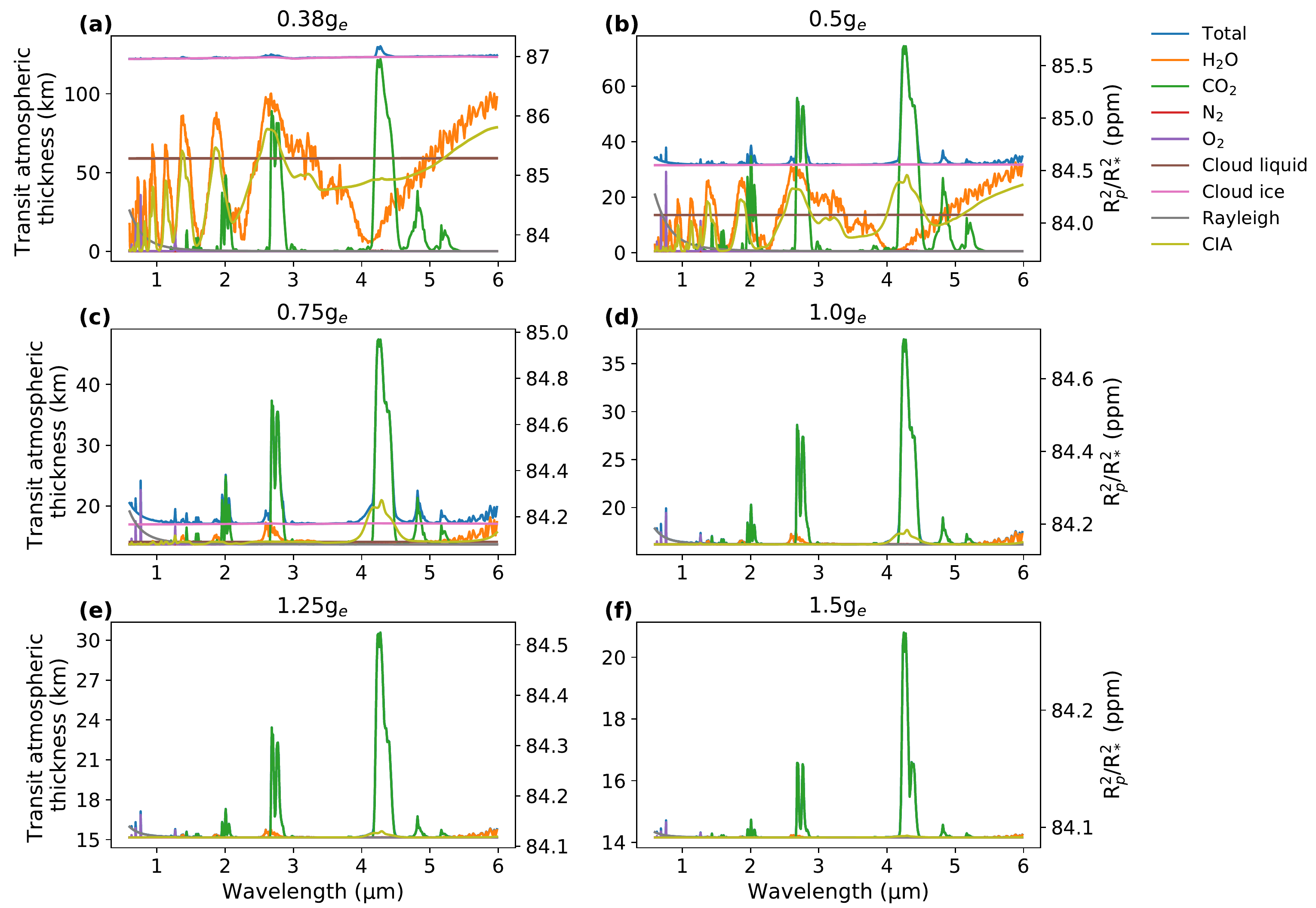}
    \epsscale{-2}
    \caption{The contribution of each component to the transmission spectra of different gravity values with a fixed stellar flux of 230 W\,m$^{-2}$. ``Rayleigh" stands for the contribution of Rayleigh scattering to the transmission spectra and CIA is short for collision-induced absorption.}
    \label{transmission_sep}
\end{figure}

To examine how gravity and stellar flux affect the characterization of the atmospheric composition of terrestrial exoplanets, we calculate the transmission spectra of selected experiments using the Planetary Spectrum Generator \citep[PSG;][]{VILLANUEVA201886}. We choose two groups of experiments. The first group is the experiments of 0.38g$_e$, 0.5g$_e$, 0.75g$_e$, 1.0g$_e$, 1.25g$_e$, and 1.5g$_e$, with a fixed stellar flux of 230 W\,m$^{-2}$. The second group is the experiments of 1.0g${_e}$, with increasing stellar flux from 230 to 290 W\,m$^{-2}$ at a 10-W\,m$^{-2}$ increment. For simplicity, the domain- and time-mean vertical profiles of the equilibrium state are used for the calculation of the transmission spectra. For the temporal variability of the transmission spectra, please see \citet{10.3389/fspas.2021.708023} and \citet{Fauchez_2022}.

We set the parent star to the Sun. The semi-major axis is 1.22 AU in the fixed stellar flux group, in which the planet can receive an annual average stellar flux of 230 W\,m$^{-2}$ at TOA. In the increasing stellar flux group, the semi-major axis is changed to make the planet receive the same amount of stellar flux at TOA as that set in the experiments.
Since planetary radius can affect the relative transit depth of the planet, we set the planet's radius to Earth's value, 6371 km, although planets of different gravities might have different radii. We calculate the transmission spectra with a resolving power of 300, from 0.6 to 6 \textmu m. The outputs of the module PSG are in the unit of transit atmospheric thickness (km). We calculated the relative transit depth (in the unit of ppm) by $(R_{p}/R_{*})^{2}$, 
in which, $R_p$ and $R_*$ are the planet's radius plus transit atmospheric thickness, and the star's radius respectively. 

The overall transit atmospheric thickness increases with decreasing gravity (Figure \ref{transmission_all}(a)). This is because clouds, especially ice clouds can extend to higher altitudes on lower-gravity planets (pink lines in Figure \ref{transmission_sep}). The absorption capability of cloud liquid water and cloud ice water is nearly constant in all wavelengths. Thus, when clouds extend to higher altitudes, the transit atmospheric thickness increases at all wavelengths. The base transit thickness contributed by cloud ice is approximately equal to the cloud top height shown in Figure \ref{profile_1}(c).

In the quasi-steady state, atmospheric components are more detectable on lower-gravity planets. In the cases of 0.5g$_e$, 0.75g$_e$, 1.0g$_e$, 1.25g$_e$, and 1.5g$_e$ with 230 W\,m$^{-2}$, convection is in the quasi-steady state. The transmission spectra are dominated by the absorption features of CO$_2$ in these cases, with two strong absorption bands centered at 4.3 and 2.7 \textmu m, and several weaker absorption bands centered at 1.6, 2.0, 4.7, and 5.2 \textmu m (Figure \ref{transmission_all}(a)). The absorption bands of O$_2$, H$_2$O, and N$_2$ can be hardly seen. As gravity decreases from 1.5g$_e$ to 0.5g$_e$, the absorption features of CO$_2$ can be seen more clearly (Figure \ref{transmission_all}(a)). This is because the atmospheric height increases with decreasing gravity, so CO$_2$ can extend and absorb stellar irradiation in higher altitudes. Although clouds mute the absorption features within the convective region, the increased atmospheric height still makes CO$_2$ more detectable on lower-gravity planets.

In the oscillatory state, high-level clouds make the atmospheric composition hard to be characterized. As convection enters the oscillatory state, cloud ice extends to very high altitudes. As seen in Figure \ref{transmission_all}(a) and \ref{transmission_sep}(a), cloud ice water nearly mutes all the spectral features in the case of 0.38g$_e$ with 230 W\,m$^{-2}$ (Figure \ref{transmission_all}(a)).

With the same gravity value, the atmospheric composition may be less detectable as the planet moves closer to the host star. As seen in Figure \ref{transmission_all}(b), the transmission spectra are dominated by the absorption bands of CO$_2$ centered at 4.3 and 2.7 \textmu m. The dominated absorption features make the atmospheric thickness reaches $\sim$42 and $\sim$33 km at 4.3 and 2.7 \textmu m respectively, and these peaks do not change much with increasing stellar flux. Meanwhile, as cloud ice water extends to higher altitudes with higher stellar flux, the base transit atmospheric thickness increases with stellar flux. These higher-level clouds mute the absorption features of CO$_2$, implying that the atmospheric composition may be less detectable with larger stellar flux.

\section{Summary and Discussions} \label{conclusions}

In this work, we have employed the cloud-resolving model SAM in the frame of RCE to investigate how gravity affects planetary climate. We are led to the following conclusions:

1. Under a fixed stellar flux, sea surface temperature increases with decreasing gravity due to both larger clear-sky greenhouse effect and stronger cloud warming effect in the small domain. The former is because lower-gravity planets have more water vapor, and the latter is because cloud water increases with decreasing gravity.

2. By increasing stellar flux, we find that under a high sea surface temperature, convection shifts from a quasi-steady state to an oscillatory state. Under the quasi-steady state, water vapor has a positive feedback, clouds have an overall negative feedback, and surface temperature increases with a nearly constant rate. During the transition to the oscillatory state, the cloud feedbacks and water vapor feedback increase substantially, resulting in rapid warming. After entering the oscillatory state, sea surface temperature increases slowly with increasing stellar flux. This is because cloud net radiative effect decreases quickly with increasing stellar flux, which indicates a trend of climate stability.

3. Because water vapor content increases with decreasing gravity, a lower-gravity planet requires a cooler climate to cause a positive radiative heating rate in the lower troposphere and to shift to the oscillatory state.

4. In the quasi-steady state, the atmospheric absorption features are more detectable on lower-gravity planets because of their larger atmospheric heights. While in the oscillatory state, the high-level clouds mute almost all the absorption features, making the atmospheric components hard to be characterized. As stellar flux increases, clouds extend to higher altitudes and makes the atmospheric absorption features less detectable.

In the quasi-steady state, sea surface temperature increases with decreasing gravity in our small-domain simulations, implying that the inner edge of the habitable zone might be further away from the host star for a lower-gravity planet. However, in the oscillatory regime, climate stability may be able to delay the climate to shift to the runaway greenhouse state in a small domain. In the oscillatory state, the climate has entered the moist greenhouse state, in which the sufficient wet stratosphere may accelerate the water vapor loss, hastening the loss of the planetary ocean.

\begin{figure}[t!]
\epsscale{1.}
\plotone{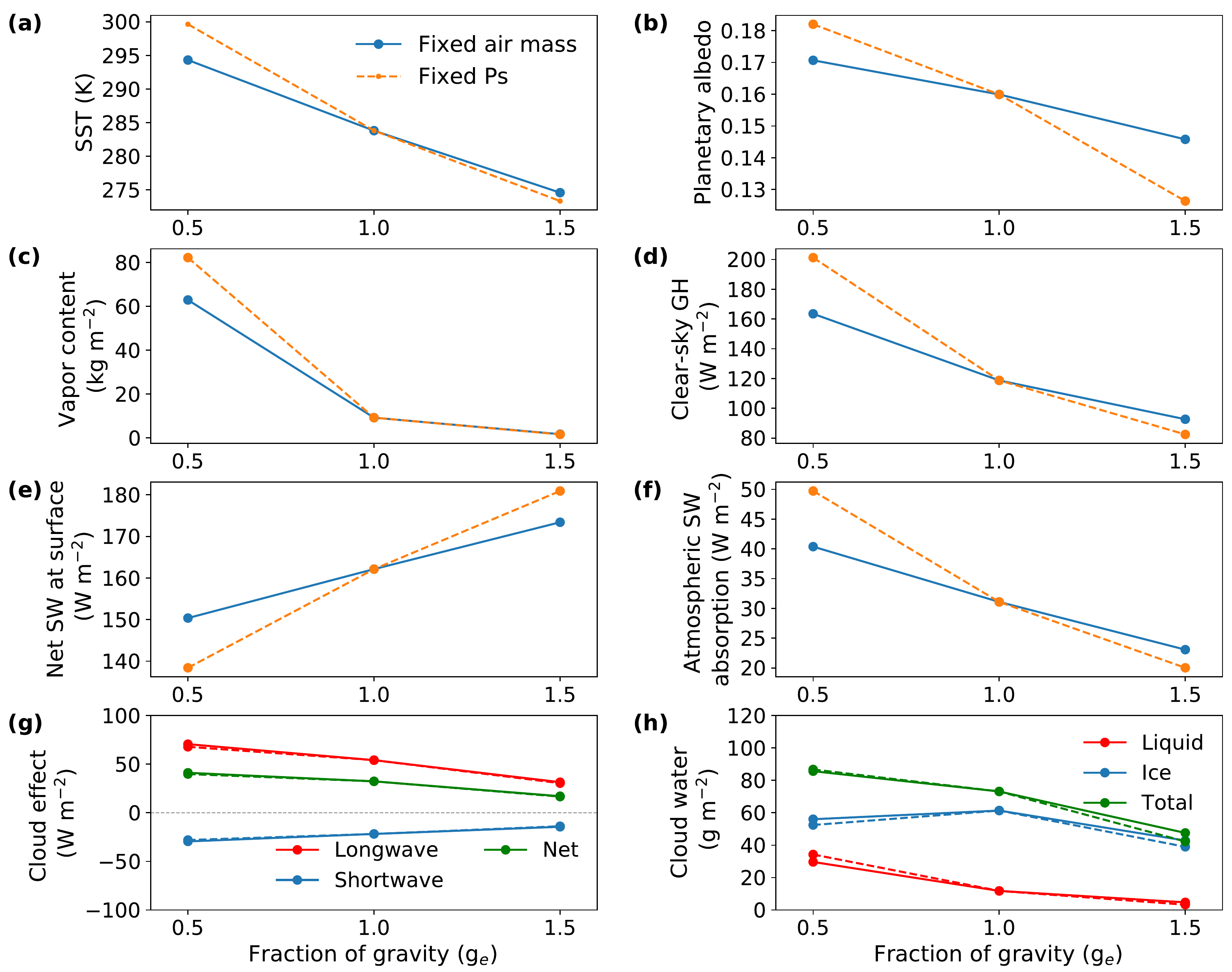}
\caption{Effects of varying planetary gravity with a fixed surface pressure (1.0 bar, orange dash lines)
and fixed air mass (1.0\,$\times$\,10$^{4}$ kg\,m$^{-2}$, blue solid lines) on sea surface temperature (a), planetary albedo (b), column water vapor content (c), the strength of clear-sky greenhouse effect (d), the net shortwave flux at the surface (e), and atmospheric shortwave absorption (f). 
Effects of varying planetary gravity on cloud longwave (red line), shortwave (blue line), and net (green line) radiative effects at TOA (g), and column cloud liquid water (red line), cloud ice water (blue line), and total cloud water (green line) (h) for experiments with a fixed surface pressure (1.0 bar, dash lines) and fixed air mass (1.0\,$\times$\,10$^{4}$ kg\,m$^{-2}$, solid lines). Noting that the case of 1.0g$_e$ in the solid and dash lines is the same. The stellar flux is 230 W\,m$^{-2}$ for all the experiments.}
\label{effect_gravity_1_1bar}
\end{figure}

\begin{figure}[t!]
\epsscale{1.}
\plotone{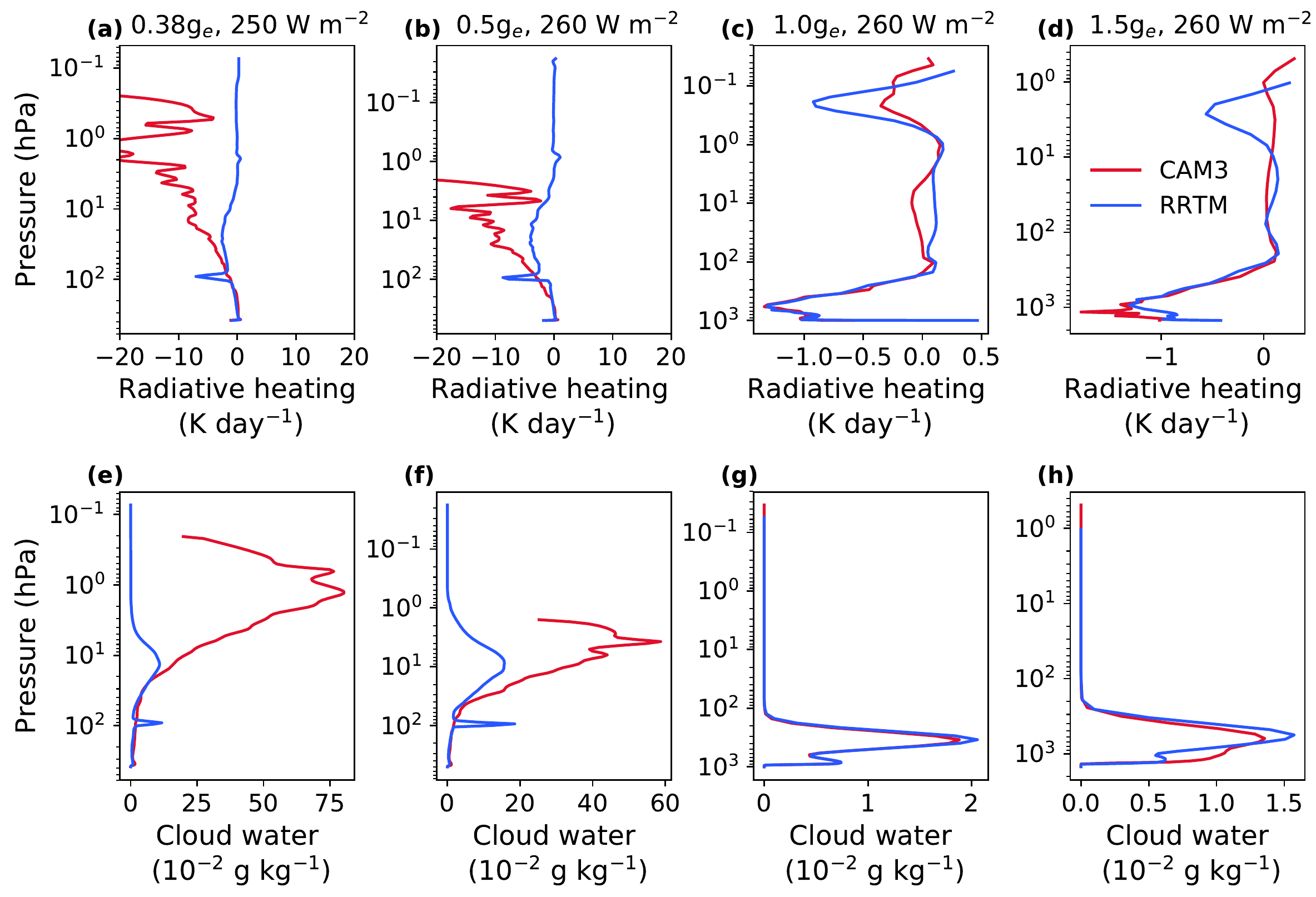}
\caption{The first row shows the comparison of radiative heating rates calculated by CAM3 (red lines) and RRTM (blue lines). The second row shows the comparison of total cloud water profiles. The cases from left to right are 0.38g$_e$ with 250 W\,m$^{-2}$, 0.5g$_e$ with 260 W\,m$^{-2}$, 1.0g$_e$ with 260 W\,m$^{-2}$, and 1.5g$_e$ with 260 W\,m$^{-2}$. The unphysical discontinuities in radiative heating rate near 100 hPa occur in the cases of 0.38g$_e$ with 250 W\,m$^{-2}$ and 0.5g$_e$ with 260 W\,m$^{-2}$, resulting in a layer of thick ``fake'' clouds near 100 hPa.}
\label{compare}
\end{figure}

We assume a fixed column air mass (mainly N$_2$) as \citet{thomson2019effects}, and the surface pressure changes proportional to gravity in our experiments. However, it is possible that the column air mass also changes with gravity. To investigate the effects of changing column air mass, we do two additional experiments, 0.5g$_e$ and 1.5g$_e$, keeping a fixed surface pressure of 1.0 bar but fixed CO$_2$ air mass, so the column air mass changes inversely proportional to gravity. For example, compared to 1.0g$_e$, the column air mass is twice in the case of 0.5g$_e$ and 2/3 in the case of 1.5g$_e$. The results are shown in Figure \ref{effect_gravity_1_1bar}. With a fixed surface pressure, sea surface temperature increases with decreasing gravity because lower-gravity planets can hold more water vapor and have more clouds, resulting in a larger greenhouse effect and cloud net radiative effect. The tendency is the same as the fixed air mass experiments discussed in Section \ref{subsec: fixed}. As air mass increases, Rayleigh scattering increases outgoing shortwave radiation to space, resulting in increased planetary albedo (Figure \ref{effect_gravity_1_1bar}(b)); multiple scattering within the atmosphere increases, and the pressure boarding effect broadens the atmospheric absorption lines, both leading to increased atmospheric shortwave absorption (Figure \ref{effect_gravity_1_1bar}(f)) and decreased net surface shortwave flux (Figure \ref{effect_gravity_1_1bar}(e)); atmospheric lapse rate increases \citep{Zhang_2020}, resulting in increased clear-sky greenhouse effect (Figure \ref{effect_gravity_1_1bar}(d)). The change in the clear-sky greenhouse effect dominates the results, leading to increased surface temperature in the case of 0.5g$_e$ and decreased surface temperature in the case of 1.5g$_e$ (Figure \ref{effect_gravity_1_1bar}(a)). Cloud water and cloud radiative effects change little with varying air mass (Figures \ref{effect_gravity_1_1bar}(g) and (h)).

We use the radiative transfer model adapted from CAM3 in this study, although it has been validated in Earth's simulation, its performance with high pressures and temperatures is not accurate enough \citep{collins2004description,collins2006formulation,yang2016differences}. It is shown that RRTM, using a correlated-k method, is more accurate than CAM3 in Earth's simulation. But we do not use RRTM in this study, because RRTM uses different look-up tables for the water vapor continuum absorption bands below and above 100 hPa, which would cause unphysical results for lower-gravity simulations. In our larger-gravity simulations, the air temperature and water vapor density near 100 hPa are low, so the discontinuous look-up tables do not cause severe problems (Figures \ref{compare}(c), (d), (g), and (h)). However, in our lower-gravity simulations, air temperature and water vapor density are large near 100 hPa, and the discontinuous water vapor look-up tables cause an unphysical radiative heating rate (Figures \ref{compare}(a) and (b)). \citet{seeley2021episodic} also found this problem. This unphysical radiative heating rate promotes a layer of thick ``fake'' clouds (Figures \ref{compare}(e) and (f)), increasing the planetary albedo dramatically.

Note that our experimental design is different from that in \citet{seeley2021episodic}.  \citet{seeley2021episodic} applied an ocean heat sink of 104.9 W\,m$^{-2}$ to mimic the atmospheric and oceanic energy export in the tropics. Since we do not know the exact value of oceanic and atmospheric heat transport on exoplanets with different gravities, we do not consider ocean heat sink in our simulations. Thus, the stellar fluxes employed in this study are much lower than that in  \citet{seeley2021episodic}. In \citet{seeley2021episodic}, the experiment with a stellar flux of 413.13 W\,m$^{-2}$ has an equilibrium sea surface temperature of about 305 K and a planetary albedo of 0.11. The net shortwave absorption of the system is equal to the net incident stellar flux minus ocean heat sink, $413.13\times(1-0.11)-104.9=262.79$ W\,m$^{-2}$. In our study, the case of 1.0g$_e$ with 270 W\,m$^{-2}$ has an equilibrium sea surface temperature of about 302 K and a planetary albedo of 0.14. The net shortwave absorption of the whole system is $270\times(1-0.14)=232.2$ W\,m$^{-2}$. Compared between these two experiments, the net shortwave absorption in our simulation is about 30 W\,m$^{-2}$ lower than that in \citet{seeley2021episodic}, but the obtained surface temperature is similar. This is because the cloud longwave radiative effect in our simulation is higher, 48.7 versus 20.0 W\,m$^{-2}$. Moreover, since the stellar flux employed in our simulation is lower, the calculated cloud shortwave radiative effect in the unit of W\,m$^{-2}$ should be weaker than that in \citet{seeley2021episodic}, even if the cloud albedo is the same. For example, the cloud shortwave radiative effect is -20.7 W\,m$^{-2}$ in our simulation and -24.6 W\,m$^{-2}$ in that of \citet{seeley2021episodic}, respectively.

The parameterizations of the microphysical scheme in SAM are based on Earth's observations, and may not be applicable to exoplanet studies \citep{lin1983bulk}. We use a single-moment microphysical scheme in this study, in which cloud particle effective radius and particle size distributions are simply prescribed \citep{khairoutdinov2003cloud}. \citet{seeley2021episodic} have demonstrated that the transition to the oscillatory state does not depend on CRMs, radiative transfers, or microphysical schemes, but simulations with more accurate radiative transfer models and microphysical schemes are still required to examine how the inner or outer edge of the habitable zone depends on planetary gravity.

Due to the limitation of computation resources, we only perform experiments in a small domain, in which large-scale circulations are not included. Large-scale circulations may influence the formation of clouds by arranging ascending and subsidence areas, water vapor distribution, and air temperature. Specifically, large-scale subsidence may reduce relative humidity and inhibit deep convection, resulting in a drier atmosphere, fewer high clouds, and a cooler climate \citep[e.g.,][]{lau1997role}. Future works including large-scale air motions are expected to verify our results.

\begin{acknowledgments}
The simulation data is available at: \href{https://doi.org/10.5281/zenodo.7197768}{https://doi.org/10.5281/zenodo.7197768}. We are grateful for the helpful discussions with Feng Ding, Daniel D.B. Koll, Xinyi Song, Mingyu Yan, Huanzhou Yang, Linjiong Zhou, Nadir Jeevanjee, Yi Huang, and Yuwei Wang. Jun Yang acknowledges support from the National Natural Science Foundation of China (NSFC) under grants 42161144011 and 42075046.
\end{acknowledgments}

\bibliography{Paper_ApJ_Jiachen}{}
\bibliographystyle{aasjournal}

\end{document}